\documentclass{aastex62}

\shorttitle{Arp 220}
\shortauthors{Wheeler et al.}

\begin{document}

\title{Arp 220: New Observational Insights into the Structure and Kinematics of the Nuclear Molecular Disks and Surrounding Gas}

\correspondingauthor{Jordan Wheeler}
\email{Wheeler1711@gmail.com}

\author[0000-0003-1678-5570]{Jordan Wheeler}
\affil{CASA University of Colorado Boulder \\
389 UCB \\
Boulder, CO 80309, USA}

\author{Jason Glenn}
\affiliation{CASA University of Colorado Boulder \\
389 UCB \\
Boulder, CO 80309, USA}
\affiliation{NASA Goddard Space Flight Center\\
Code 665, 8800 Greenbelt Road\\
Greenbelt, MD 20771}

\author{Naseem Rangwala}
\affiliation{NASA Ames Research Center \\
Space Science \& Astrobiology Division \\
MS 245-6, Bldg 245, Rm 107F\\
Moffet Field, CA 94035, USA}

\author{Adalyn Fyhrie}
\affiliation{CASA University of Colorado Boulder \\
389 UCB \\
Boulder, CO 80309, USA}

\begin{abstract}

ALMA cycle 3 observations of $^{12}$CO $ J = 3\rightarrow2$, $^{13}$CO $J =  4\rightarrow 3$, SiO J = $8 \rightarrow 7$, and HCN J = $ 5 \rightarrow 4$ are presented. Significant extended emission is detected in $^{12}$CO J $ = 3\rightarrow2$ with a morphology that is indicative of m = 2 tidal features, suggesting gas inflow. In addition, outflow for both nuclei are found in the $^{12}$CO J $ = 3\rightarrow2$. Significant SiO absorption is detected in the western nucleus. HCN that is morphologically distinct from CO is detected in both nuclei.
These observations are compared to non-LTE radiative transfer models created using the Line Modeling Engine (LIME) for simple gas dynamics to gain insight into how physical parameters,
such as rotational velocity, turbulent velocity, gas temperature, dust temperature, and gas mass can reproduce the observed kinematic and spatial features. 
The eastern nucleus is found to be best modeled with an inclusion of a temperature asymmetry from one side of the disk to the other.
It is also found that the western nucleus is optically thick even in the less abundant species of $^{13}$CO, absorbing significant amounts of continuum radiation.

\end{abstract}


%%%%%%%%%%%%%%%%%%%%%%%%%%%%%%%%%%%%%%%%%%%%%%%%%%%%%%%%%%%%%%%%%%%%%%%%%%%%%%%
%                        Introduction                                         %
%%%%%%%%%%%%%%%%%%%%%%%%%%%%%%%%%%%%%%%%%%%%%%%%%%%%%%%%%%%%%%%%%%%%%%%%%%%%%%%

\section{Introduction} \label{sec:intro}

Arp 220 is an ultra-luminous infrared galaxy (ULIRG) in the late stages of a merger between two gas-rich galaxies. 
The Arp 220 molecular gas system consists of two counter rotating nuclear disks separated by 1", $\sim 400$ pc, embedded in a larger rotating extended emission disk \citep{Sakamoto1999}.
At a distance of only 77 Mpc, ALMA's high spatial resolution resolves the emission from the two merging nuclei and provides insight into their gas dynamics in this extreme environment.
Arp 220 has very large nuclear star formation rates with $\Sigma_{SFR} = 10^{3.7}$ M$_{\odot}$ yr$^{-1}$ kpc$^{-2}$ for the eastern nucleus and $\Sigma_{SFR} = 10^{4.1}$ M$_{\odot}$ yr$^{-1}$ kpc$^{-2}$ for the western nucleus \citep{Barcos2015}.
It shows an unprecedentedly large dust optical depth with $\tau \sim 1$ at $\lambda = 2.6$ mm in the western nucleus \citep{Scoville2017}. 
\citet{Scoville2017} also found that the kinematics of the western nucleus in CO J = $1 \rightarrow 0$ suggest a central Keplerian mass of $\sim8 \times 10^{10}$ M$_\odot$, indicating the potential presence of a super-massive black hole.
Unambiguous identification or exclusion of the presence of an active galactic nuclei, AGN, in the Arp 220 system remains elusive (e.g. \citep{Sakamoto2017}).
\citet{Barcos2015} where unable to provide evidence that the radio emission at 33 GHz is dominated by production from an AGN.
Yet, the extreme emission of the Arp 220 system in continuum measurements hint at the potential for a AGN, or at the least, a very extreme starburst \citep{Wilson2014}. Observations using $\gamma$-rays also suggest the need for an AGN power source \citep{Yoast2017}, and the large column densities observed in OH$^+$, H$_2$O$^+$, and H$_2$O additionally point to a requirement of an X-ray luminous source such as an AGN with a luminosity of 10$^{44}$ erg/s \citep{Rangwala2011,Gonzalez2012,Gonzalez2013}.

Not only is dust obscuration in the Arp 220 system significant, but molecular absorption features further obscure the center of the disks \citep{Rangwala2015,Tunnard2015,Aalto2015,Martin2016,Scoville2017}. In addition, Arp 220 has evidence for molecular outflows from several lines \citep{Sakamoto2009,Sakamoto2017,Barcos2018}. 
Specifically, \citet{Barcos2018} found evidence of a fast strongly collimated outflow in the western nucleus with a gas depletion time scale of order 10 Myr.
Resolved measurements of Arp 220 have revealed complex structure and a wealth of new information that was previously not realizable with single dish measurements.
In this way, Arp 220 provides the opportunity to understand how physical conditions such as morphology and kinematics can inform the integrated flux density measurements for sources that cannot be resolved at high redshift.

This paper first presents the ALMA observations and data reduction in Section 2. Following this, in Sections 3, 4, and 5, the implications of the observations that can be directly observed from the data are discussed.
In particular, in Section 4 and 5 the extended emission morphology and outflowing molecular gas are examined in detail. 
Finally, in Section 6 kinematic non-LTE radiative models for both the eastern and western nuclei are presented and compared with the observations. 
The implications of the physical parameters derived from these models for the nuclear environments are then considered.

\section{Observations and Data Reduction} \label{sec:data}

This paper presents data from two sets of observations: (centered on) $^{12}$CO J = $3\rightarrow 2$ and $^{13}$CO J = $4 \rightarrow 3$. 
Arp 220 was observed in CO J = $3 \rightarrow 2$ with ALMA's band 7 receivers on August 17$^{th}$ of 2016 (Project 2015.1.00736.S) during cycle-3 in a configuration with baselines ranging from 15 m to 1.4 km. 
The bands observed were centered on 350.6 GHz (continuum), 352.4 GHz (continuum), 340.6 GHz ($^{12}$CO J = $3\rightarrow 2$), and 342.4 GHz (SiO J = $8 \rightarrow 7$).
The total on-source integration time was 4 minutes using 38 antennas. 
The correlator was configured in low-resolution wide-bandwidth TDM mode to give a spectral resolution of 15.625 MHz. During the observation, the weather was excellent with a median PWV of 0.32 mm. The beam size was 0.244" x 0.175" with a position angle of 0.34$^\circ$. 
The 1~$\sigma$ noise in a cleaned 20 km/s spectral channel is $\sim$1.5 mJy/beam.

Arp 220 was observed in $^{13}$CO J = $4 \rightarrow 3$ with ALMA's band 8 receivers on July 29$^{th}$ and 30$^{th}$ of 2016 as well as on August 17$^{th}$ and 18$^{th}$ of 2016 (Project 2015.1.00736.S) during cycle-3 in a configuration with baselines ranging from 15 m to 1.4 km. 
The bands observed were centered on 432.0 GHz ($^{13}$CO J = $4 \rightarrow 3$), 433.8 GHz (HCN J = $5 \rightarrow 4$/continuum), and 421.8 GHz (continuum).
The total on-source integration time was 102 minutes using 51 antennas. 
The correlator was configured in low-resolution wide-bandwidth TDM mode to give a spectral resolution of 15.625 MHz. The median PWV for the observations was 0.65 mm. 
The beam size was 0.177" x 0.125" with a position angle of -11.7$^\circ$. 
The 1~$\sigma$ noise in a cleaned 20 km/s spectral channel is $\sim$1.1 mJy/beam.

Imaging and cleaning were carried out using the Common Astronomy Software Applications package CASA \citep{CASA}. 
The observations were flagged, calibrated, and imaged by ALMA staff prior to delivery. 
Additional calibration and imaging were performed to verify the quality of the reduced observations. 
The delivered observations were found to be of high quality and are thus presented here without additional processing.
All observations have been corrected for primary beam attenuation.
The observations were imaged using Briggs weighting, with robust set to 0.5, resulting in a balance between point source and extended flux density sensitivity. 
Both observation sets were continuum subtracted using CASA's uvcontsub routine.
However, for the non-LTE models discussed in Section \ref{sec:LIME}, images were produced rather than UV data.
In this case, CASA's imcontsub routine was used for continuum subtraction rather than uvcontsub. 
To confirm that the different continuum subtraction routines do not produce different results, the observations were reanalyzed using both uvcontsub and imcontsub.
The continuum subtracted observations were then compared and it was confirmed that the two routines produce the same line profiles within the noise levels of the observations. 

%%%%%%%%%%%%%%%%%%%%%%%%%%%%%%%%%%%%%%%%%%%%%%%%%%%%%%%%%%%%%%%%%%%%%%%%%%%%%
%               Observations                                                %
%%%%%%%%%%%%%%%%%%%%%%%%%%%%%%%%%%%%%%%%%%%%%%%%%%%%%%%%%%%%%%%%%%%%%%%%%%%%%

\section{Observed Morphology and Kinematics} \label{sec:observations}

\begin{figure}[ht!]
\plotone{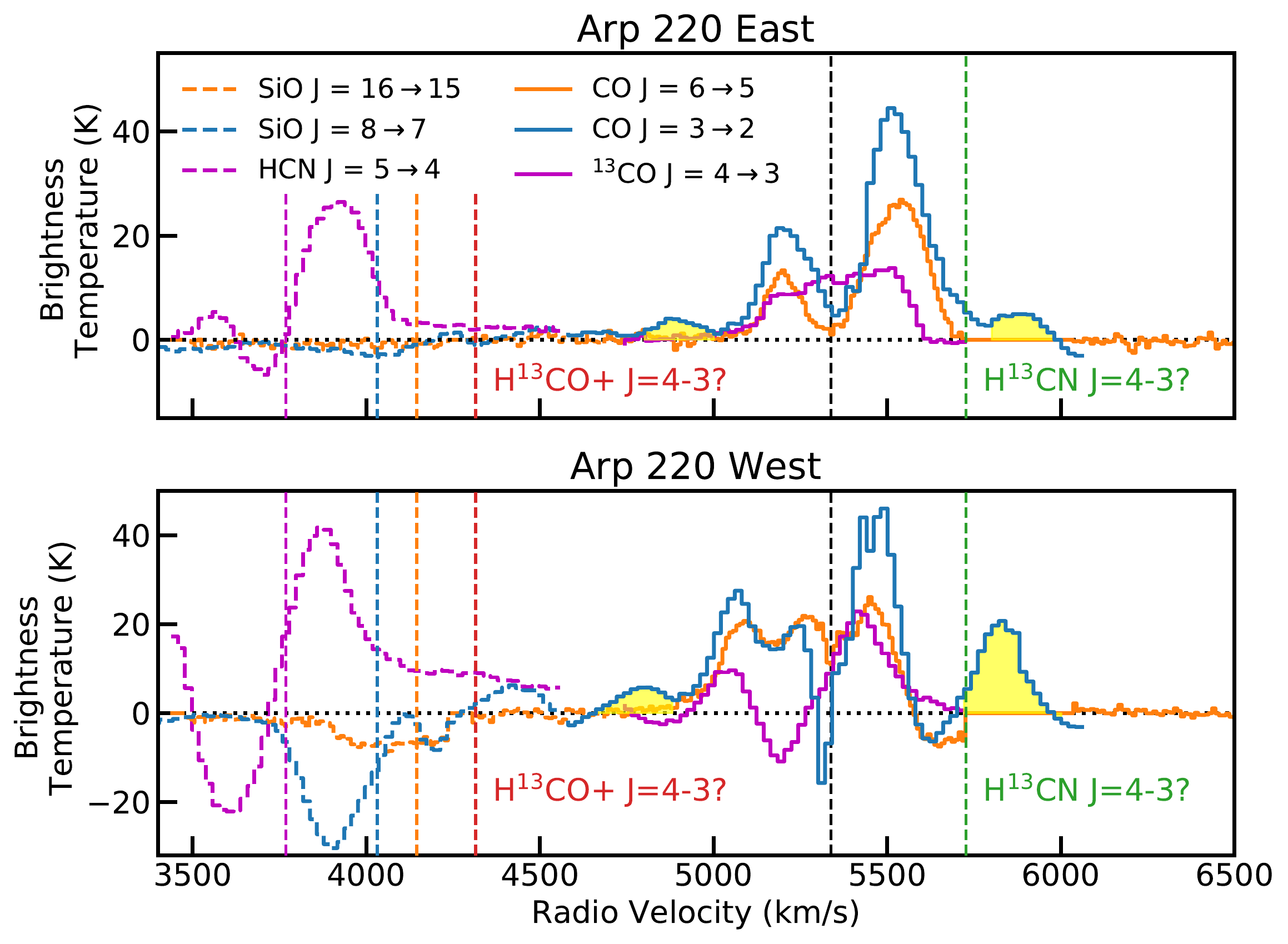}
\caption{Continuum subtracted CO line profiles in the two nuclei. Note the deep absorption features at the centers of the eastern nucleus and western nucleus. High-velocity outflows seen in $^{12}$CO J = $3 \rightarrow 2$ are highlighted in yellow. The line profiles were extracted at the center of the continuum emission at 427.8 GHz (i.e 15:34:57.292, +23.30.11.33 and 15:34:57.223, +23.30.11.49) for the eastern and western nuclei respectively. This is because each fitted continuum center location is different by 1 to 2 pixels with the 427.8 GHz continuum having the median center location of the 3 data sets. Assumed rest frequency of $^{12}$CO J = $3 \rightarrow 2$ is 345.796 GHz and $^{13}$CO J = $4 \rightarrow 3$ is 440.765 GHz. $^{12}$CO J = $6 \rightarrow 5$ observations are from \citet{Rangwala2015}. For $^{12}$CO J = $3\rightarrow 2$ 1 Jy/beam = 248 K, for $^{13}$CO J = $4\rightarrow 3$ 1 Jy/beam = 294.5 K, and for $^{12}$CO J = $6\rightarrow 5$ 1 Jy/beam = 30 K. The  systemic velocity of the CO lines are shown as vertical dashed black lines. The systemic velocity of the other lines are shown in corresponding colored vertical dashed lines. In addition, the systemic velocity of the H$^{13}$CN J = 4 $\rightarrow$3 line that may contaminate some of the CO J = 3 $\rightarrow$ 2 emission and the H$^{13}$CO+ J = 4 $\rightarrow$3 line that may contaminate some of the SiO J = 8 $\rightarrow$ 7 absorption are shown.} \label{fig:lines}
\end{figure}

The overall morphology is similar to that described by \citet{Sakamoto1999}: two small nuclear disks embedded in a larger extended emission disk. 
The extended emission disk is well detected in $^{12}$CO J = $3 \rightarrow 2$ but is not in $^{13}$CO J = $4 \rightarrow 3$. 
Continuum emission morphology for the nuclei is similar in both bands after accounting for the different beam sizes at 365.5 and 427.8 GHz. 
The continuum flux densities and their Gaussian fits are listed in Table \ref{meas_table}. The line profiles for the molecular species at the centers of each of the nuclei are shown in Figure \ref{fig:lines}.

\subsection{$^{12}$CO J = $3 \rightarrow 2$}

\begin{figure}[ht!]
\plotone{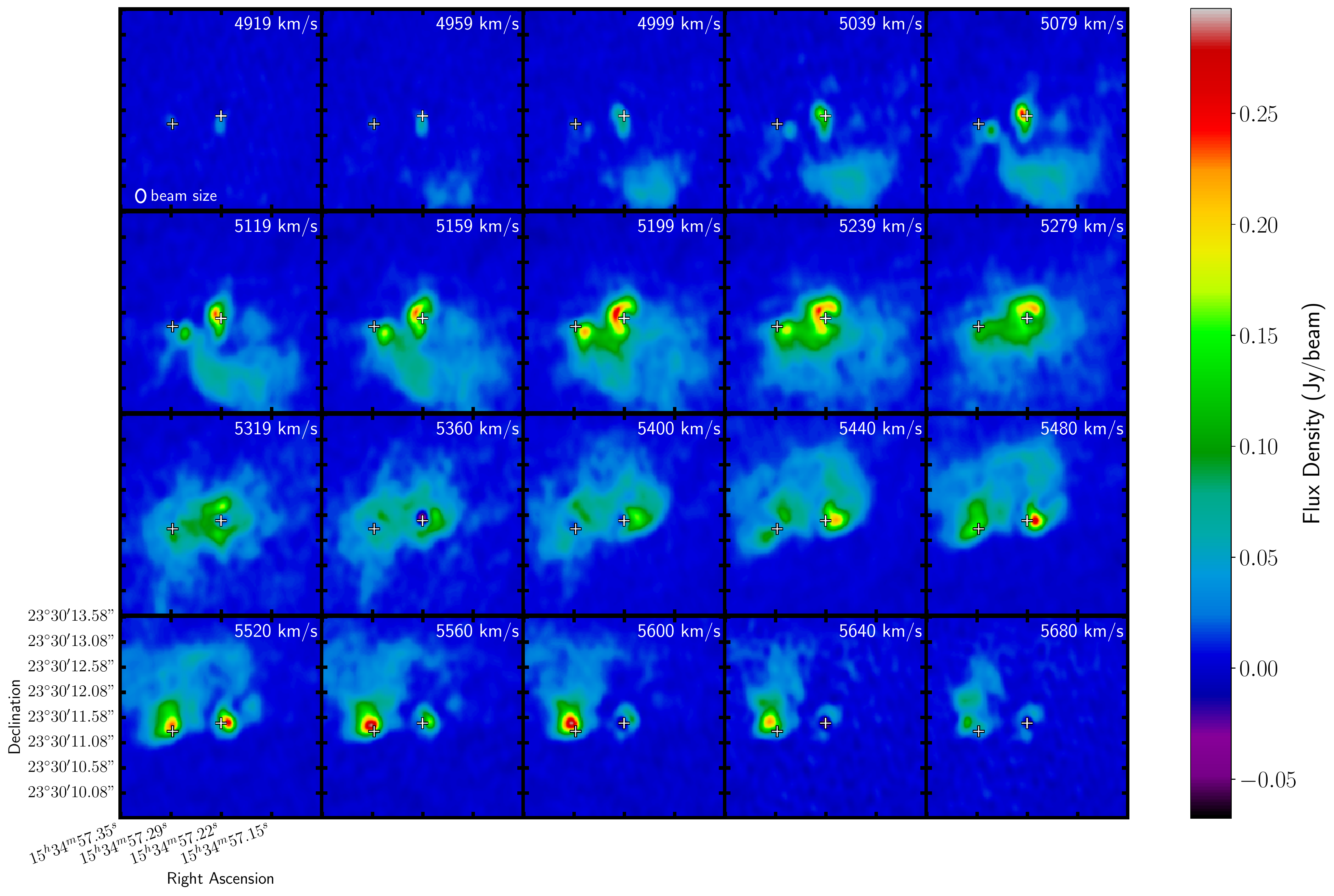}
\caption{Spectrometer channel maps for $^{12}$CO J = $3 \rightarrow 2$. Radio velocity for each panel is shown in white text. Crosses indicate the center of continuum emission in each nuclei. There is a large negative dip (below the continuum level) at 5360 km/s and 5640 km/s. The velocity structure of the extended disk structure of Arp 220 is well resolved. \label{fig:12COchan}}
\end{figure}

The spectrometer channel maps for $^{12}$CO  J = $3 \rightarrow 2$ are shown in Figure \ref{fig:12COchan}, and the nuclear velocity kinematic profiles are shown in Figure \ref{fig:lines}.
%{\color{red} [If you refer to Fig. 5 here, you'll have to make it Fig. 2.]}. 
For $^{12}$CO  J = $3 \rightarrow 2$, emission can be seen starting at a velocity of $\sim$4,750 km/s in both nuclei.
At 4,959 km/s the blue extended emission becomes detectable in the southwest.
Progressing from low velocities to high velocities, the emission spirals in toward the eastern nucleus.
The extended emission then propagates out from the western nucleus, and the most redshifted emission disappears north of the eastern nucleus.
The strong emission of the western nucleus is evident starting at 5,039 km/s and becomes redder toward the west to a velocity of 5,560 km/s.
At 5,360 km/s and 5,640 km/s there is a strong absorption dip.
The eastern nucleus emission changes from blue in the southwest to red in the northeast, ranging in velocities from 5,159 km/s to 5,640 km/s. 

$^{12}$CO J = $3 \rightarrow 2$ moment 0, 1, and 2 maps are shown in Figure \ref{fig:12COmom}. 
Moments 1 and 2 were constructed utilizing only pixels with a signal greater than 4 times the rms of pixels without signal.
Moment 0 clearly shows a large extended emission complex surrounding both nuclei.
This extended emission emphasized by the 8 Jy/beam km/s contour in Figure \ref{fig:12COmom} has a size of 3.5 arcseconds, corresponding to a physical size of 1.3 kpc (1 arcsecond = 373 pc).
The eastern and western nuclei moment 0 emission is highlighted with a 45 Jy/beam km/s contour. 
The eastern nucleus has a physical size along its major axis of 0.6 arcseconds, or 220 pc. The western nucleus has a size of 0.5 arcseconds or 180 pc in the east-west direction and 0.7 arcseconds, or 270 pc, in the north-south direction. 
Moment 0 also shows an asymmetric brightness for the eastern nucleus with increased flux density for the northeast part of this disk. 
The central line profile, Figure \ref{fig:lines}, and line profiles throughout the nuclear disk also show an asymmetric brightness with the red emission exhibiting enhanced brightness compared to the blue. 
For the western nucleus, moment 0 shows a diamond-shaped core with a significant reduction in flux density at the center. 
%Moment 1 shows the velocity gradient for the extended emission and the eastern nucleus is from southwest to northeast and for the western nucleus from west to east.
Moment 1 shows the velocity gradient for the extended emission and the nuclei, indicating that the velocity gradient for the extended emission and the eastern nucleus is from southwest to northeast and for the western nucleus it is from west to east.
Moment 2 shows that the highest velocity dispersion is located along the rotational axis of the eastern and western nuclei and has peaks of 200 and 300 km/s for each respectively. 
The extended emission has a velocity dispersion of 100 to 150 km/s.

\subsection{SiO J = $8 \rightarrow 7$}

\begin{figure}[ht!]
\plotone{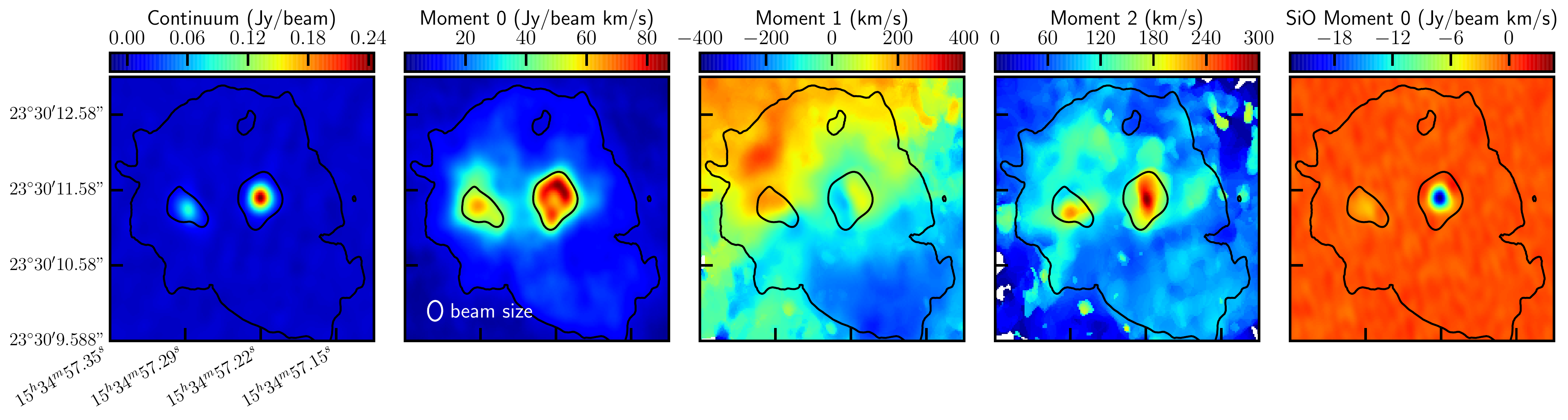}
\caption{346.5 GHz continuum emission, moments 0, 1, and 2 for $^{12}$CO J = $3 \rightarrow 2$, and moment 0 of Si0 J = $8 \rightarrow 7$. Moment 0 contours of 8, 45 Jy/beam  km/s of $^{12}$CO J = $3 \rightarrow 2$ are shown in black for all images to highlight the nuclei and extended emission. A systemic velocity of 5338 km/s is assumed for moment 1.  \label{fig:12COmom}}
\end{figure}

Moment 0 of SiO J = $8 \rightarrow 7$ is shown in Figure \ref{fig:12COmom}. 
The SiO J = $8 \rightarrow 7$ absorption in the western nucleus is significant, but only just detectable in the eastern nucleus.
The absorption in Figure \ref{fig:12COmom} is mostly unresolved and coincident with the center of the western nucleus.
The absorption in the western nucleus has two distinct velocity components: a very deep blue component and a less so red component. % think this is suppose to be a colon
The blue component shows a FWHM velocity of 240 km/s and the red component of only 70 km/s.
If this absorption is tracing shocked gas in an outflow in this nucleus, it would be expected that the blue component, which would be in the foreground, potentially would exhibit more absorption than the red component, which is occurring in the background of the disk-emitting mass. 
However, the velocities may be tracing the underlying velocity structure of the rotating disk rather than outflow velocities. 
Potentially the  blue absorption of SiO could be contaminated by possible emission coming from the H$^{13}$CO$^+$ J = $4 \rightarrow 3$ line depicted in Figure \ref{fig:lines}.
SiO J = $16 \rightarrow 15$ has also been seen in absorption in the western nucleus \citep{Rangwala2015} and the J = $5 \rightarrow 4$ line has been seen in emission \citep{Martin2011}.

Si is known to be depleted in the gas phase throughout the interstellar medium \citep{Haris2016}. 
This is due to Si's high condensation temperature of $\sim$1300 K which causes it to be one of the first elements to be incorporated into silicate dust grains in the atmospheres of AGB stars \citep{Savage1996}. 
As a result, if Si is observed in the gas phase, a mechanism to extract the Si from silicate dust grains is required. 
Shocks are the most common mechanism invoked for the destruction of dust grains in interstellar mediums \citep{Draine1979}.
Thus,  it is generally accepted that observations of Si in the gas phase are an indicator of the presence of shocks.
In particular, \citet{Schilke1997} showed that a shock velocity of 25 km/s and densities of 10$^5$ cm$^{-3}$ is sufficient to extract the Si from dust grains creating observable SiO in the gas phase.
The presence of SiO absorption features indicates the presence of such shocks within the western nucleus. 
In the eastern nucleus, a SiO absorption feature is also detectable but the absorption is much less significant. 
This indicates that shocks, while present in the eastern nucleus, are much less significant than in the western nucleus.

\subsection{$^{13}$CO J = $4 \rightarrow 3$ }
\begin{figure}[ht!]
\plotone{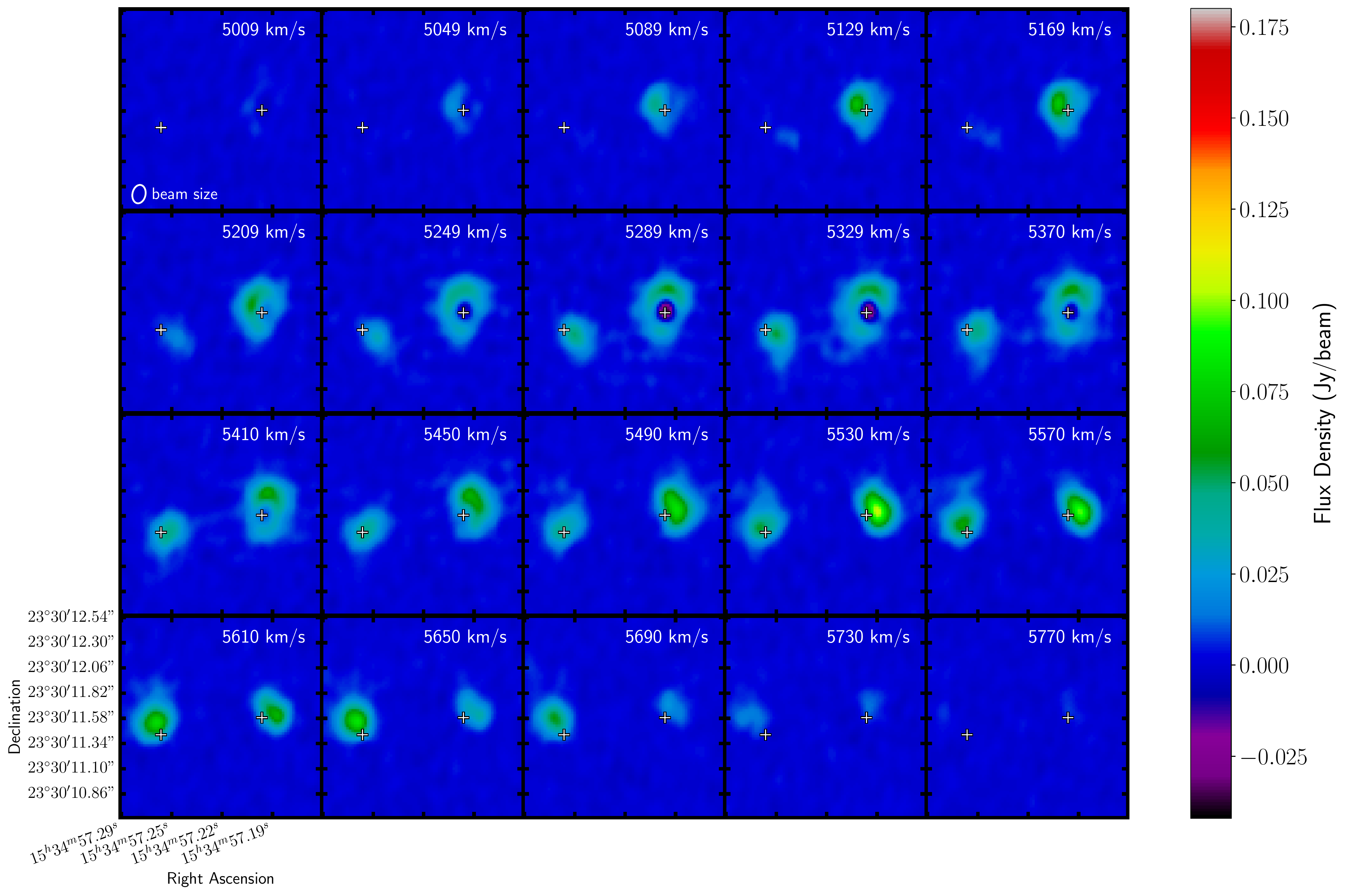}
\caption{Channel maps for $^{13}$CO J = $4\rightarrow 3$. Radio velocity for each panel is shown in white text. Crosses indicate the center of continuum emission in each nucleus.  \label{fig:13COchan}}
\end{figure}

\begin{figure}[ht!]
\plotone{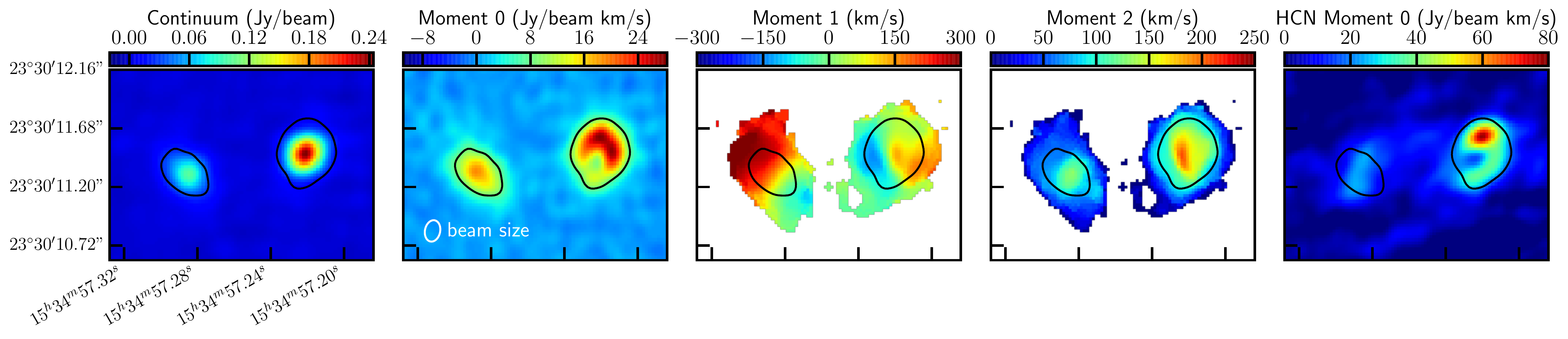}
\caption{427.8 GHz continuum emission, moments 0, 1, and 2 for $^{13}$CO J = $4 \rightarrow 3$, and moment 0 of HCN J = $5 \rightarrow 4$. A moment 0 contour of 8 Jy/beam km/s  of $^{13}$CO J = $4 \rightarrow 3$ } is shown in black in all images to highlight the nuclei. A systemic velocity of 5338 km/s is assumed for moment 1. \label{fig:13COmom}
\end{figure}

The spectrometer channel maps for $^{13}$CO  J = $4 \rightarrow 3$ are shown in Figure \ref{fig:13COchan}, and the nuclear velocity kinematic profiles are shown in Figure \ref{fig:lines}.
$^{13}$CO J = $4 \rightarrow 3$ moment 0, 1, and 2 maps are shown in Figure \ref{fig:13COmom}. 
The $^{13}$CO J = $4 \rightarrow 3$ observations do not detect the extended emission disk but do resolve the two nuclei with high S/N.
The $^{13}$CO observations are similar to the $^{12}$CO observations for the nuclear disks with a few notable exceptions.
For the eastern nucleus, the continuum subtracted line profiles do not exhibit an absorption feature at zero velocity.
The profiles are still asymmetric from red to blue as in the $^{12}$CO J = $3 \rightarrow 2$ observations, but much less so.
These two properties, along with the modeling in Section \ref{sec:LIME}, help to inform the possible sources for this asymmetry.
For the western nucleus, similar kinematics as in the $^{12}$CO J = $3 \rightarrow 2$ observations are seen, including a large central absorption feature.
However, the absorption in $^{13}$CO J = $4 \rightarrow 3$ is broader than and shifted in velocity relative to the absorption in $^{12}$CO J = $3 \rightarrow 2$.
In addition, the western nucleus shows a faint high-velocity red wing indicating a potential outflow.

The $^{12}$CO J = $3 \rightarrow 2$ to $^{13}$CO J = $4 \rightarrow 3$ ratio was examined by convolving the $^{13}$CO beam to the larger size of the $^{12}$CO beam and then dividing each pixel in each spectral channel. 
The pixels examined were limited to S/N $>$ 3 in $^{13}$CO and the frequency difference between the different transitions, {6:8}, was compensated for. 
For most of the pixels the ratio is small, $<$ 10, suggesting that at least the $^{12}$CO is optically thick in both nuclei.
A small number of pixels located in the gaseous bridge between the two nuclei at -80 km/s were found to have a ratio of approximately 40. This suggests a normal abundance ratio between the two species for at least some of the gas. It should be noted that these are close but different transitions which may reflect slightly different gas components, i.e. cold versus warm gas.

%%%%%%%%%%%%%%%%%%%%%%%%%%%%%%%%%%%%%%%%%%%%%%%%%%%%%%%%%%%%%%%%%%%%%%%%%%%%
%                        HCN
%%%%%%%%%%%%%%%%%%%%%%%%%%%%%%%%%%%%%%%%%%%%%%%%%%%%%%%%%%%%%%%%%%%%%%%%%%%%

\subsection{HCN J = $5 \rightarrow 4$}
Bright HCN J = $5 \rightarrow 4$ emission is seen in both nuclei. 
Table \ref{meas_table} contains the measured properties, Figure \ref{fig:lines} the line profiles, and Figure \ref{fig:13COmom} the moment 0 map.
The morphology of the HCN emission has some striking dissimilarities when compared to the CO emission.
This is most noticeable in the eastern nucleus where the emission is almost perpendicular in the position angle to the CO emission. 
Also, in the western nucleus, the emission is most concentrated in the poles of the disk rather than in the plane of the disk.
In both nuclei, the HCN feature clearly dips below the zero continuum level.
This indicates that the HCN is absorbing some of the dust emission.
In the eastern nucleus even more pronounced asymmetry from blue to red in the central line profiles is seen in HCN J = $5 \rightarrow 4$ emission, compared to the CO observations.
In the western nucleus this may also be the case but the blue emission is cut off as it approaches the edge of our observation band.
This cutoff also results in the measured flux density for the western nucleus in Table \ref{meas_table} under-measuring the total flux density in HCN for this nucleus.

HCN, because of its large dipole moment, requires a high collision rate to maintain a substantial population above the ground state. 
Thus, it is often thought of as a tracer of dense gas. 
However, it can also be vibrationally excited through radiative pumping from 14 $\mu$m continuum radiation fields \citep{Aalto2007,Sakamoto2010}.
In order to distinguish between the excitation methods, HCN can be compared with other species, such as HNC and HCO$^+$, or the vibrationally excited $v_2 = 1$ states can be observed directly \citep{Aalto2007,Sakamoto2010,Martin2016}.
The observations in this paper do not probe HNC or HCO$^+$.
The vibrationally excited HCN J = $5 \rightarrow 4$ $\nu_2 = 1$, $\ell$ = 1e line, is only +35 km/s from the $\nu = 0$ line, making it indistinguishable from the ground state. 
In addition, the $\ell$ = 1f line is at -1,400 km/s, outside our observation band. 
Since we do not have access to the direct or indirect observations that would allow us to distinguish between excitation methods, we do not attempt to draw any conclusions about the excitation method. 
However, \citet{Martin2016} observed the $\nu_2 = 1$, $\ell$ = 1f line and used it to predict the contribution of the radiatively pumped $\nu_2 = 1$, $\ell$ = 1e to the $\nu = 0$ line in both HCN J = $4 \rightarrow 3$  and J = $3 \rightarrow 2$.
They found that there is a significant but not dominant contribution for those lines (see Figure 3 \citep{Martin2016}) that also causes increased red emission. 
In addition, the fact that the HCN line is as bright as the CO emission may indicate at least some radiative pumping.

HCN emission is potentially seen perpendicular to the disk because 14 $\mu$m photons from hot dust are more likely to escape out of the less optically thick poles orthogonal to the disk rather than through the plane of the disk. 
These photons can then be absorbed and re-emitted in the submillimeter by the HCN molecules.
In addition, the HCN in the western nucleus shows a strong central absorption dip below the continuum level.
This indicates that the HCN is absorbing the continuum emission, which reduces its apparent flux density at the centers of the nuclei, Figure \ref{fig:lines}. 
The apparent flux density reduction also contributes to its polar appearance. 
A similar preferentially polar morphology is seen the vibrationally excited HCN emission in IC 860 \citep{Aalto2019}.
This type of morphology is explained by different optical depths of the continuum and line emission for the radiatively pumped line emission \citep{Gonzalez2019}.
It could be concluded that the HCN is preferentially tracing the outflowing gas more than the gas in the disk.
However, the majority of the gas does not have velocities with large deviations from the systemic velocity that would indicate an outflow.

\subsection{CO morphology versus rotational transition}

\begin{figure}[ht!]
\plotone{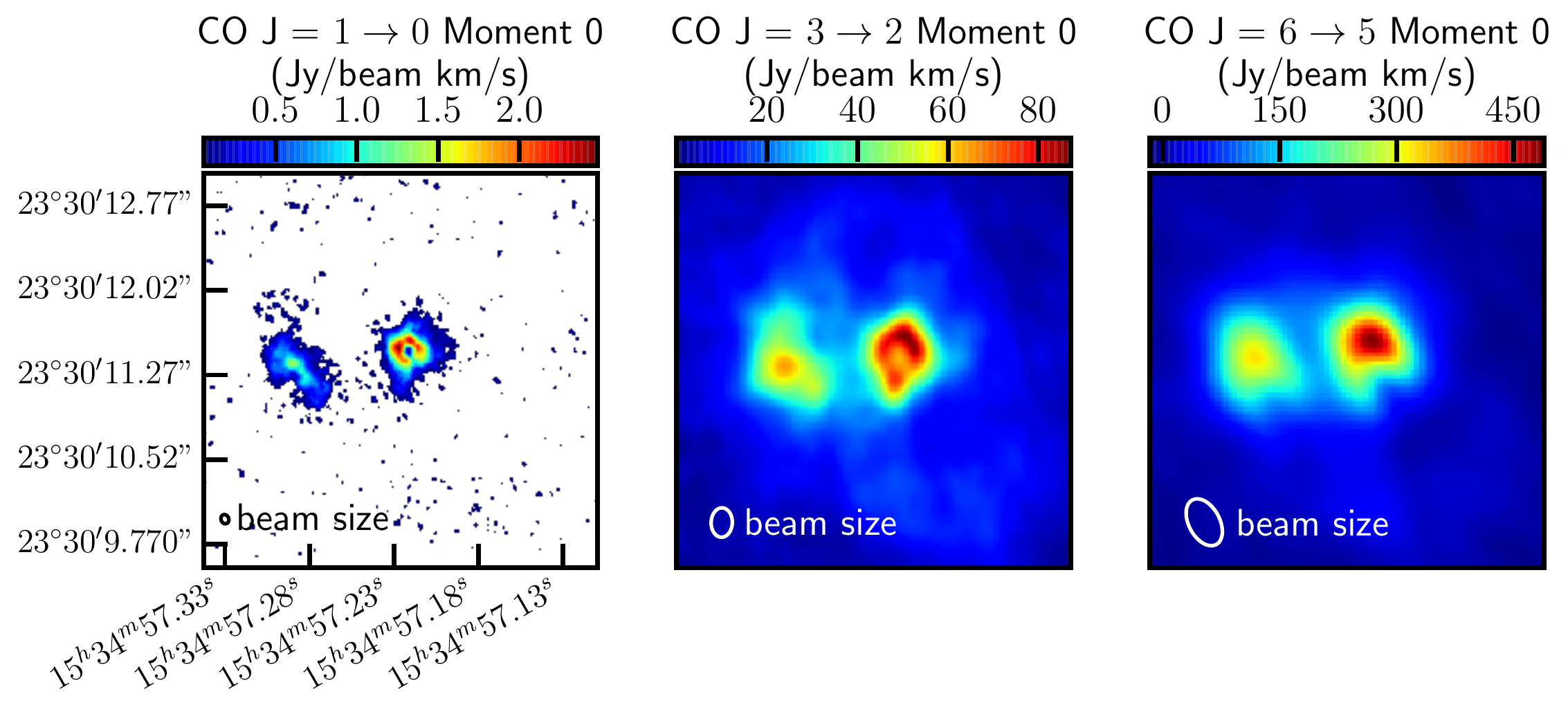}
\caption{Moments 0 for $^{12}$CO J = $1 \rightarrow 0$, $^{12}$CO J = $3 \rightarrow 2$, and $^{12}$CO J = $6 \rightarrow 5$.  \label{fig:mom_compare}}
\end{figure}

The moment 0 maps of  $^{12}$CO J = $1 \rightarrow 0$, $^{12}$CO J = $3 \rightarrow 2$, and $^{12}$CO J = $6 \rightarrow 5$ are shown in Figure \ref{fig:mom_compare}. 
The $^{12}$CO J = $1 \rightarrow 0$ data are that of \citet{Scoville2017}, but imaged with robust = 0.5 rather than 1, and a velocity resolution of 20 km/s rather than 40 km/s. Moment 0 of $^{12}$CO J = $1 \rightarrow 0$ was produced by only using pixels with signal greater than 4 times the rms of noise within the channels. 
The observed structure for each transition is similar between the transitions even given the very different beam sizes. 
One noticeable distinction is that the asymmetry in brightness from one side of the nuclear disks to the other in each of the two nuclei becomes more pronounced with observations of higher J$_{upper}$ transitions. 
Such that in J = $6 \rightarrow 5$ the north eastern part of the eastern nucleus is much brighter than the southwestern side. 
Likewise, for the western nucleus the north and western parts of the disk appear brighter than the rest of the disk. 
This suggests that hotter gas traced by the higher J transitions is not as evenly distributed throughout the nuclear disks.

%\startlongtable
\begin{longrotatetable}
\begin{deluxetable}{c|cccc}
\tablecaption{Arp 220 Continuum and Line Measurements \label{meas_table}}
\tablehead{
\colhead{Quantity} & \colhead{Eastern Nucleus} & \colhead{Western Nucleus}  & \colhead{Extended Emission} & \colhead{Units}}
\startdata
\hline
\multicolumn{5}{c}{346.5 GHz Continuum (Beam 0.23" x 0.16" PA = -0.3$^\circ$)} \\
\hline
Peak flux location & (15:34:57.294, +23.30.11.34) & (15:34:57.224, +23.30.11.50) & Not detected& J2000 \\
Integrated flux density (1"x1") & 220 & 440 & ... & mJy  \\
Peak flux density & $80.1 \pm 2.8 $ & $237 \pm 6.4$ & ... & Jy/beam\\
Gaussian FWHM & 0.367 x 0.257, PA 26.9$^\circ$ $\pm$ 3.8$^\circ$ &  0.265 x 0.245, PA 70$^\circ$ $\pm$ 14$^\circ$ & ... & arcseconds \\
Gaussian fit integrated flux density & 195.4 $\pm$ 7.6& 398 $\pm$ 16& ... & mJy \\
Deconvolved FWHM & 0.298 x 0.171, PA 40.7$^\circ$ $\pm$ 5.6$^\circ$ & 0.205 x 0.076, PA 85$^\circ$ $\pm$ 14$^\circ$  & ... & arcseconds\\
\hline
\multicolumn{5}{c}{427.8 GHz Continuum (Beam 0.18" x 0.13" PA = -23.2$^\circ$)} \\
\hline
Peak flux location & (15:34:57.292, +23.30.11.33) & (15:34:57.223, +23.30.11.49) & Not detected& J2000 \\
Integrated flux density (1"x1") & 380 & 680 & ... & mJy  \\
Peak flux density & $93 \pm 2.4$ & $237 \pm 3$ & ... & Jy/beam\\
Gaussian FWHM & 0.329 x 0.262, PA 43.2$^\circ$ $\pm$ 4.3$^\circ$ &  0.268 x 0.234, PA 139.7$^\circ$ $\pm$ 3.7$^\circ$ & ... & arcseconds \\
Gaussian fit integrated flux density & 342 $\pm$ 11& 636 $\pm$ 11& ... & mJy \\
Deconvolved FWHM & 0.299 x 0.194, PA 49.9$^\circ$ $\pm$ 4.4$^\circ$ & 0.208 x 0.183, PA 105$^\circ$ $\pm$ 12$^\circ$  & ... & arcseconds\\
\hline
\multicolumn{5}{c}{$^{12}CO J = 3 \rightarrow 2$ (Beam 0.23" x 0.17" PA = -0.34$^\circ$)} \\
\hline
Peak flux location & (15:34:57.29, +23.30.11.41) & (15:34:57.23, +23.30.11.49) & Not Gaussian & J2000 \\
Integrated flux density (1"x1") & 700 & 800 & 1800 (3.5"x3.5") & Jy km/s \\
Peak flux density & $54.7 \pm 2.8 $ & $80.3 \pm 4.7 $ & Not Gaussian & Jy/beam km/s\\
Gaussian FWHM & 1.018 x 0.79, PA 30.8$^\circ$ $\pm$ 8.1$^\circ$ &  0.781 x 0.663, PA 156$^\circ$ $\pm$ 14$^\circ$ & Not Gaussian & arcseconds \\
Gaussian fit integrated flux density & 1037 $\pm$ 56& 977 $\pm$ 62& Not Gaussian & Jy km/s \\
Deconvolved FWHM & 0.99 x 0.77 PA 32.7$^\circ$ $\pm$ 9.8$^\circ$ & 0.745 x 0.636, PA 152$^\circ$ $\pm$ 28$^\circ$  & Not Gaussian & arcseconds\\
 $L^\prime_{CO_{3 \rightarrow2}}$ & $1.1 \times 10^9$ & $1.25 \times 10^9$ & $2.8 \times 10^9$ & K km/s pc$^2$ \\
$L_{CO_{3 \rightarrow2}}$ & 1.5 $\times$ 10$^6$ & 1.7 $\times$ 10$^6$ & 3.9 $\times$ 10$^6$ & L$_\odot$ \\
$L_{CO_{3 \rightarrow2}}/L_{FIR}$ $^a$ & $30^{+45}_{-21.5} \times 10^{-6}$ &  $2.6^{+18.4}_{-0.9} \times 10^{-6}$ & ... & ... \\
\hline
\multicolumn{5}{c}{$^{13}CO J = 4 \rightarrow 3$ (Beam 0.18" x 0.12" PA = -11.7$^\circ$)} \\
\hline
Peak flux location & (15:34:57.29, +23.30.11.34) & (15:34:57.22, +23.30.11.54) & Not Detected  & J2000 \\
Integrated flux density (1"x1") & 110 & 180 & ... & mJy km/s \\
Peak flux density & 17.86 $\pm$0.6 & 26.3 $\pm$ 1.7 & ... &Jy/beam km/s \\
Gaussian FWHM & 0.427 x 0.285 PA 44.4$^\circ$ & 0.408 x 0.368 PA 3.5$^\circ$ $^a$ & ... & arcseconds \\
Gaussian fit integrated flux density & 98.1 $\pm$ 3.8 & 179 $\pm$ 13 & ... & Jy km/s \\
Deconvolved FWHM &0.403 x 0.233, PA 48.3$^\circ$ & 0.371 x 0.343 PA 15$^\circ$ $^b$ & ... & arcseconds \\
$L_{^{13}CO_{4\rightarrow3}}$ & 290 & 490  & ... & L$_\odot$ \\
$L_{^{13}CO_{4\rightarrow3}}/L_{FIR}$ $^a$ & $5.8^{+9.2}_{-4} \times 10^{-9}$ & $7.8^{+54.2}_{-2.9} \times 10^{-10}$ & ... & ... \\
\hline
\multicolumn{5}{c}{SiO $J = 8 \rightarrow 7$ (Beam 0.24" x 0.18" PA = -1.59$^\circ$)} \\
\hline
Peak flux location & (15:34:57.30,+23.30.11.36) & (15:34:57. +23,23.30.11.49) & Not Detected  & J2000 \\
Integrated flux density (1"x1") & -13 & -26 & ... & Jy km/s \\
Peak flux density& -28.7 $\pm$ 0.4 & -23.0 $\pm$ 0.9 & ... &Jy/beam km/s \\
Gaussian FWHM & 0.431 x 0.381, PA 51$^\circ$ & 0.237 x 0.289 PA 12.1$^\circ$ & ...& arcseconds \\
Gaussian fit integrated flux density & -11.0 $\pm$ 1.8 & 24.0 $\pm$ 0.9 & ... & Jy km/s \\
Deconvolved FWHM & 0.385 x 0.304, PA 66$^\circ$ & Not resolved & ... & arcseconds \\
\hline
\multicolumn{5}{c}{HCN $J = 5 \rightarrow 4$ (Beam 0.18" x 0.13" PA = -26.3$^\circ$) } \\
\hline
Peak flux location & (15:34:57.29,+23.30.11.39) & (15:34:57.22 +23,23.30.11.53) & Not Detected  & J2000 \\
Integrated flux density (1"x1") & 240 & 490 & ... & Jy km/s \\
Peak flux density & 21.1 $\pm$ 2.0 & 47.4 $\pm$ 5.2 & ... &Jy/beam km/s \\
Gaussian FWHM & 0.707 x 0.398, PA 126$^\circ$ & 0.576 x 0.426, PA 141$^\circ$ & ... & arcseconds \\
Gaussian fit integrated flux density & 248 $\pm$ 25 & 488 $\pm$ 59 & ... & Jy km/s \\
Deconvolved FWHM & 0.685 x 0.371, PA 125$^\circ$ & 0.547 x 0.406, PA 139$^\circ$ & ... & arcseconds \\
\enddata
\tablenotetext{a}{Using the $L_{FIR}$ measured from the ALMA 691 GHz continuum emission from \citet{Wilson2014}. The large range in uncertainty in this ratio is from the large systematic uncertainty on the $L_{FIR}$.}
\tablenotetext{b}{The position angle on $^{13}$CO western nucleus is highly uncertain and the profile is not well described by a Gaussian.}
\tablecomments{Table showing the measured parameters from the observations. All uncertainties are 1 $\sigma$.
}
\end{deluxetable}
\end{longrotatetable}

%%%%%%%%%%%%%%%%%%%%%%%%%%%%%%%%%%%%%%%%%%%%%%%%%%%%%%%%%%%%%%%%%%%%%%%%%%%%%%%%%%%%
%                EXTENDED EMISSION
%%%%%%%%%%%%%%%%%%%%%%%%%%%%%%%%%%%%%%%%%%%%%%%%%%%%%%%%%%%%%%%%%%%%%%%%%%%%%%%%%%%%

\section{Arp 220 Extended Emission}
Due to the outstanding S/N in the $^{12}$CO $3 \rightarrow$ 2 observations, low-level signal revealing extensive structure in the extended emission surrounding the two nuclei was detected.
Significant flux density is seen extending over 3.5". 
The highest velocity blue gas is located far away from the two nuclei in the southwest; whereas, the highest velocity red gas is located just slightly north of the eastern nucleus. 
Of particular interest, the total emission in this extended region is greater than the total emission of the two nuclei combined, Table \ref{meas_table}. 
The extended emission in J = $3 \rightarrow 2$ comprises more than 50\% of the total flux density; whereas, in the J = $6 \rightarrow 5$ observations from \citet{Rangwala2015} it is responsible for less than 25\%. 
This decrease in flux density ratio at high J indicates that the extended gas is cooler and/or less dense than the gas in the disks.
The fractional difference between what components dominate the total emission for different transitions indicates that for single-dish observations different transitions may be probing different physical gas components rather than just different temperature gas.

\begin{figure}[ht!]
\plotone{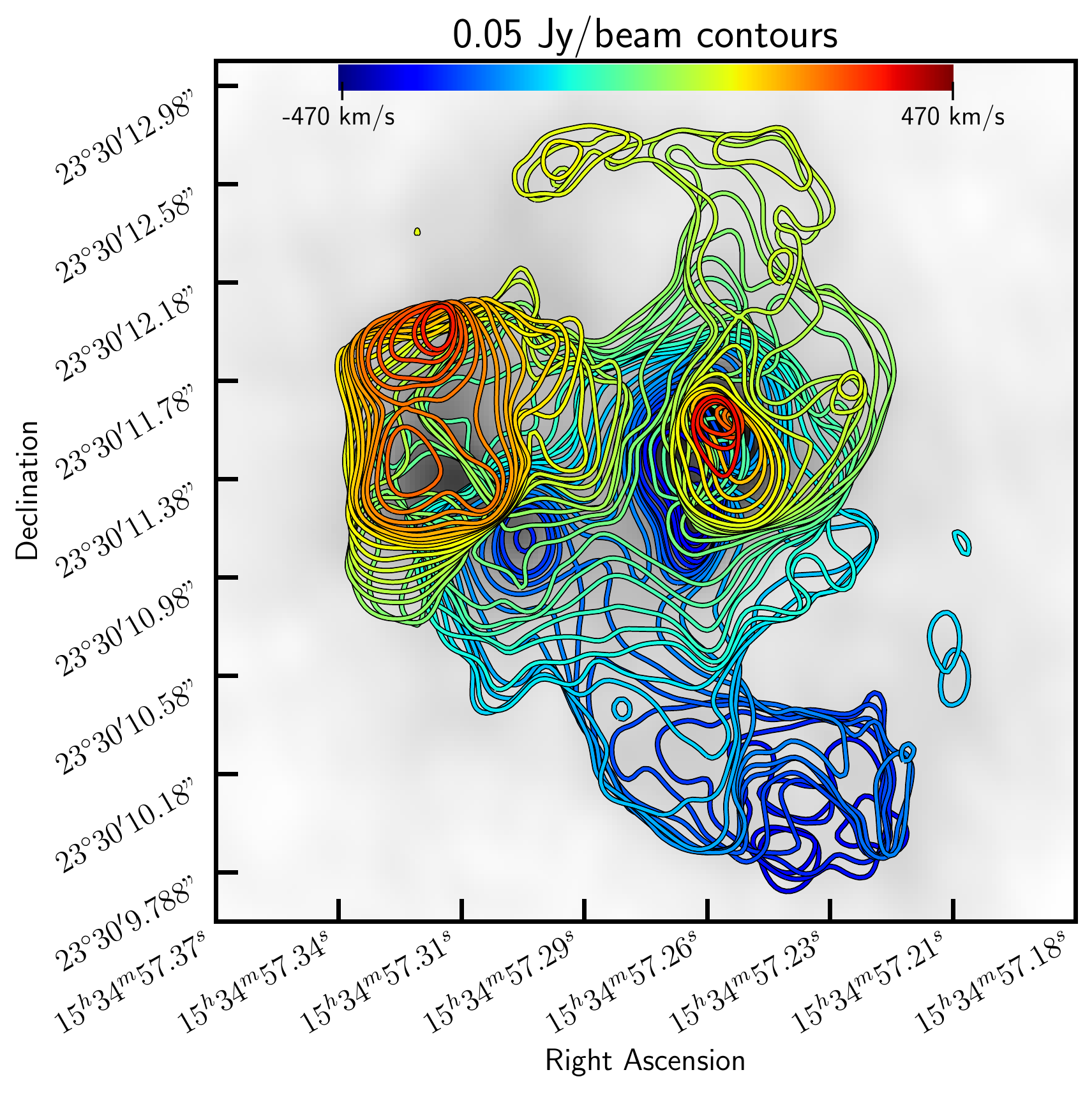}
\caption{$^{12}$CO $3 \rightarrow$ 2 0.05 Jy/beam contours are plotted for each velocity bin from -470km/s to +470km/s. This contour value highlights structure that resembles tidal tails. Moment 0 map is shown in grey scale as the background.\label{fig:extended}}
\end{figure}

If a 0.05 Jy/beam contour level is chosen at each velocity, structure is revealed, Figure \ref{fig:extended}.
It should be noted, however, if a smaller contour level is taken, i.e. 0.03  Jy/beam, this structure is much harder to discern and the red arm shown in Figure \ref{fig:extended} merges with the highest velocity red gas north of the eastern nucleus.
The structure seen with 0.05 Jy/beam contours is reminiscent of tidal tails caused by the merging of two galaxies such as those predicted by \citet{Cox2006,Johnston2008,Teyssier2010,Sparre2016}.
In this case, a m = 2 spiral asymmetry is observed. 
Tidal features are most commonly observed using stars as the tracer medium, but in some cases have been observed in CO \citep{Taylor2001,Espada2012}. 
In this case, compared to more common observations of tidal features as well as the size scales of the models cited above, these features are quite close to the nuclear disks, no more than a 1 kpc away. 
The FWHM of these features is 200 km/s on the blue side and 140 km/s on the red side. This m = 2 asymmetry could be the result of tidal torques acting on the gas.
If so, these torques could provide the mechanism to remove angular momentum of the surrounding gas allowing the gas to fall inward \citep{Barnes1991} providing the necessary fuel for the intense observed nuclear star formation.

%%%%%%%%%%%%%%%%%%%%%%%%%%%%%%%%%%%%%%%%%%%%%%%%%%%%%%%%%%%%%%%%%%%%%%%%%%
%                         outflow                                        %
%%%%%%%%%%%%%%%%%%%%%%%%%%%%%%%%%%%%%%%%%%%%%%%%%%%%%%%%%%%%%%%%%%%%%%%%%%

\section{$^{12}$CO J = $3\rightarrow2$ outflow}
\begin{figure}[ht!]
\plotone{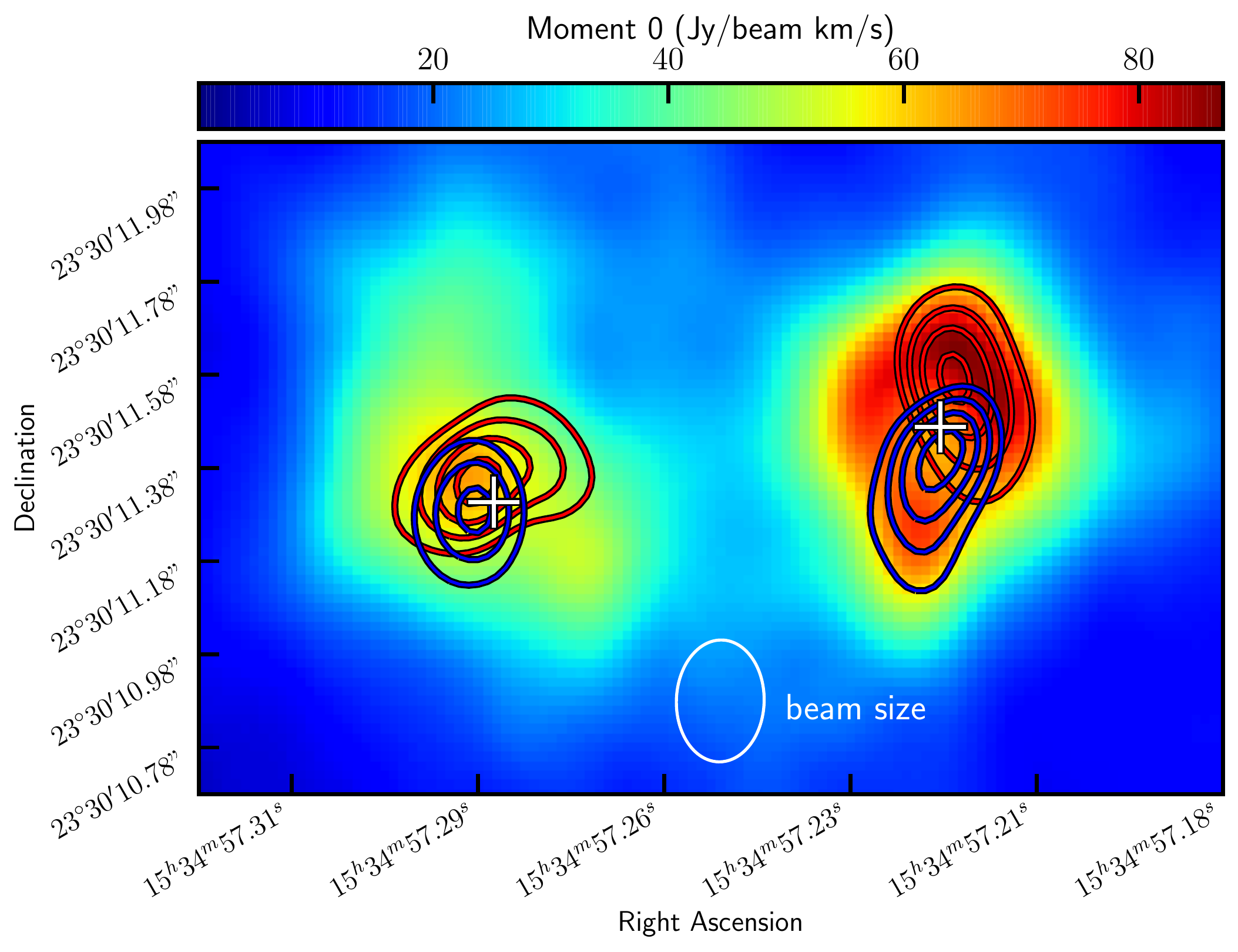}
\caption{$^{12}$CO J - $3\rightarrow2$ red and blue high-velocity gas moment 0 maps tracing outflowing gas are shown as contours. This is overlaid on a $^{12}$CO J - $3\rightarrow2$ moment 0 map that includes gas at all velocities. The velocities used to make the contours are highlighted in yellow in Figure \ref{fig:lines}.  The eastern nucleus has 1.5, 2.25, 3.0, and 3.75 Jy/beam km/s contours and the western nucleus has 1.0, 1.75, 2.5, 3.25  Jy/beam km/s contours for the blue outflow and has 6.0, 9.0, 12.0, 15.0 Jy/beam km/s contours for the red outflow. Crosses indicate the center of continuum emission in each nucleus. \label{fig:outflow}}
\end{figure}

In the $^{12}$CO J - 3-2 observations, significant outflow is detected in both the eastern and western nuclei, Figure \ref{fig:outflow}.
Significant detection of collimated outflows in the western nucleus have been previously observed; however, this marks the first highly significant detection of a collimated outflow in the eastern nucleus. 
The outflow shown in Figure \ref{fig:outflow} was constructed by making a moment 0 map for both the red and blue outflows using the high-velocity gas emission highlighted in yellow in Figure \ref{fig:lines}. 
The orientation and velocity of the outflow detected in the western nucleus match the outflow detected in HCN J = $1 \rightarrow 0$ and CO J = $1 \rightarrow 0$ by \citet{Barcos2018}. 
While an examination of Figure \ref{fig:outflow} shows a red outflow that seems to merge with the red portion of the disk, it should be noted that the velocities used to measure the outflow are distinct (much higher velocity) from what seems to be disk related velocities in CO. 
However, for the red outflow at a velocity of 5700 km/s some of the emission could be from the H$^{13}$CN J = $4 \rightarrow 3$ line. 
With this in mind, it is likely that the red outflow is contaminated by H$^{13}$CN emission. 
In particular, if there is red emission from H$^{13}$CN that is associated with the disk, this would explain why the red outflow seems to merge with the red side of the disk in Figure \ref{fig:outflow}.

The western nucleus outflow has a total luminosity for the red component of 32 Jy km/s and the blue component of 3.4 Jy km/s. 
We can estimate the mass in these outflows by using the conversion factor 0.8 M$_\odot$ (K km/s pc$^2$)$^{-1}$ \citep{DownesSolomon1998}. 
This conversion factor is for $^{12}$CO J = $1 \rightarrow 0$.
To convert from CO J = $3 \rightarrow 2$ to CO J = $1 \rightarrow 0$ we use a scale factor that is the ratio of the CO line luminosity for each of the eastern and western nuclei respectively from  CO J = $3 \rightarrow 2$ to CO J = $1 \rightarrow 0$.
This brightness temperature ratio may be different, however, in the outflowing gas, especially if the outflowing gas has different optical depths compared to the nuclei as a whole. 
The flux density is converted to a line luminosity using the equation,
\begin{equation}
L^\prime_{CO} =3.25 \times 10^7 \times I \times \nu_{obs}^{-2} \times D_L^2 \times (1 + z)^{-3} \textrm{,}
\end{equation}
where I is in Jy km/s, $\nu_{obs}$ is in GHz, and $D_L$ is in Mpc \citep{Solomon1992}.
To compute the line luminosity scale factors, the measured flux density values from \citet{Scoville2017} for CO J = $1 \rightarrow 0$ of 47.3  Jy km/s and 27.8 Jy km/s are used to compute the CO J = $ 1 \rightarrow 0$ line luminosities.
These values along with $\nu_{obs} = 117.4$ GHz yields $L^\prime_{CO} = 6.3 \times 10^8 $ and $3.7 \times 10^8$ K km/s pc$^2$.
$L^\prime_{CO} $ for  CO J = $3 \rightarrow 2$ for the eastern and western nuclei are given in Table \ref{meas_table}.
Together, this yields a conversion factor of 1/2 for the western nucleus and 1/3 for the eastern nucleus.
Using this equation a mass of 2.0$\times$10$^7$ M$_\odot$ for the red outflow and 2.0$\times$10$^6$ M$_\odot$ for the blue outflow in the western nucleus is derived. This result is 4 times smaller than the blue outflow mass found in CO J = $1\rightarrow 0$ by \citet{Barcos2018} of 8$\times$10$^6$ M$_\odot$; yet, the red outflow mass is much more than was previously found, 2.5$\times$10$^6$ M$_\odot$. 
Again, the red mass estimate is potentially contaminated by H$^{13}$CN emission so it is likely an overestimate of the true mass. The sizes of the outflow were fit using CASA and are shown in Table \ref{out_table}. 
Using these sizes and the outflow average velocities, both corrected for viewing angle based on the derived inclination angles (Table \ref{east_table}, \ref{west_table}), an age is calculated by dividing the corrected size by the corrected velocity. Using this age, the mass loss rates of 50 and 8 M$_\odot$ yr$^{-1}$ for the red and blue outflows respectively were obtained. 
For the blue outflow, this is somewhat smaller than the results of \citet{Barcos2018} of 35 M$_\odot$ yr$^{-1}$. 
For the red outflow, 3 times more mass than the \citet{Barcos2018} value of 12 M$_\odot$ yr$^{-1}$ was found, possibly from contamination from H$^{13}$CN emission. 

Similarly, for the eastern nucleus, the outflow has a total luminosity for the red component of 8.6 Jy and the blue component of 6.4 Jy. This yields a mass of 5$\times$10$^6$ M$_\odot$ and 4$\times$10$^6$ M$_\odot$ for the red and blue outflows respectively. 
For the eastern nucleus, the outflows are not well resolved so the sizes could be smaller than those fitted.
In this case, the sizes in Table \ref{out_table} are taken as upper limits to the true sizes. 
Using these sizes lower limits for the mass loss rates are found to be 19 M$_\odot$ yr$^{-1}$ for both the red and blue outflow respectively. Given masses of the eastern and western nuclei of $\sim 2 \times 10^9$ M$_\odot$, these mass loss rates yield a depletion time scale of order 10 Myr.

\startlongtable
\begin{deluxetable}{c|cccc}
\tablecaption{Arp 220 Outflow Measurements\label{out_table}}
\tablehead{
\colhead{Quantity} & \colhead{Eastern Nucleus Red} & \colhead{Eastern Nucleus Blue}  &\colhead{Western Nucleus Red$^{b}$}  & \colhead{Western Nucleus Blue}}
\startdata
\hline
Velocity (km/s) & 450-610 & -(390-590) & 390-650 & -(470-670)\\
Flux density (Jy km/s)  & 8.6 & 6.4 & 32 & 3.4 \\
$L^\prime_{CO 3 \rightarrow 2} $ (K km/s pc$^2$) & 1.3$\times$10$^7$ & 1.0$\times$10$^7$ & 5.0$\times$10$^7$ & 5.0$\times$10$^6$ \\
$L^\prime_{CO 1 \rightarrow 0} $ $^a$ (K km/s pc$^2$) & 4.3$\times$10$^5$ & 3.3$\times$10$^6$ & 2.5$\times$10$^7$ & 2.5$\times$10$^6$ \\
M$_{H2}^{b,c}$ (M$_\odot$) & 3.3$\times$10$^6$ & 2.7$\times$10$^6$ & $<$ 2.0$\times$10$^7$ & 2.0$\times$10$^6$ \\
Size  (pc x pc) & $<$ 129 x 24 & $< $ 96 x 79 & 121 x 50 & 90 x 50\\
Age  (yr) & $< 1. \times 10^5$ & $< $ $1.4\times10^5$ & $4.0\times10^5$ & $2.5\times10^5$\\
$\dot{M}$ (M$_\odot$ yr$^{-1}$) & $>$ 19 & $> $ 19 & $<$ 50 & 8\\
%\hline
%\multicolumn{5}{c}{HCN $J = 5 \rightarrow 4$} \\
%\hline
\enddata
\tablenotetext{a}{Derived by scaling $L^\prime_{CO 3 \rightarrow 2}$ by 1/2 and 1/3 for the western and eastern nuclei respectively.}
\tablenotetext{b}{Calculated using 0.8 M$_\odot$ (K km/s pc$^2$)$^{-1}$. }
\tablenotetext{c}{For the red outflow at a velocity of 5700 km/s  some of the emission could be from the H$^{13}$CN J =$4 \rightarrow 3$ line.  With this in mind, it is likely that the red outflow is contaminated by H$^{13}$CN emission resulting in an overestimate of the red outflow mass, especially for the western nucleus, where optical depths are larger.}
\end{deluxetable}

%blue 340 GHz red 338GHz

\section{LIME modeling}\label{sec:LIME}
LIME, the Line Modeling Engine \citep{Brinch2010}, is a non-LTE spectral line radiation transfer code for 3D models in arbitrary geometries, \url{https://github.com/lime-rt/lime}.
LIME carries out radiative transfer in 3 dimensions. After the computationally expensive radiative transfer, the user can then choose any viewing angle from which to create an image. 
This enables simultaneous calculation of several different inclination angles of the disk when searching for a representative model of the observations.  

\begin{figure}[ht!]
\plotone{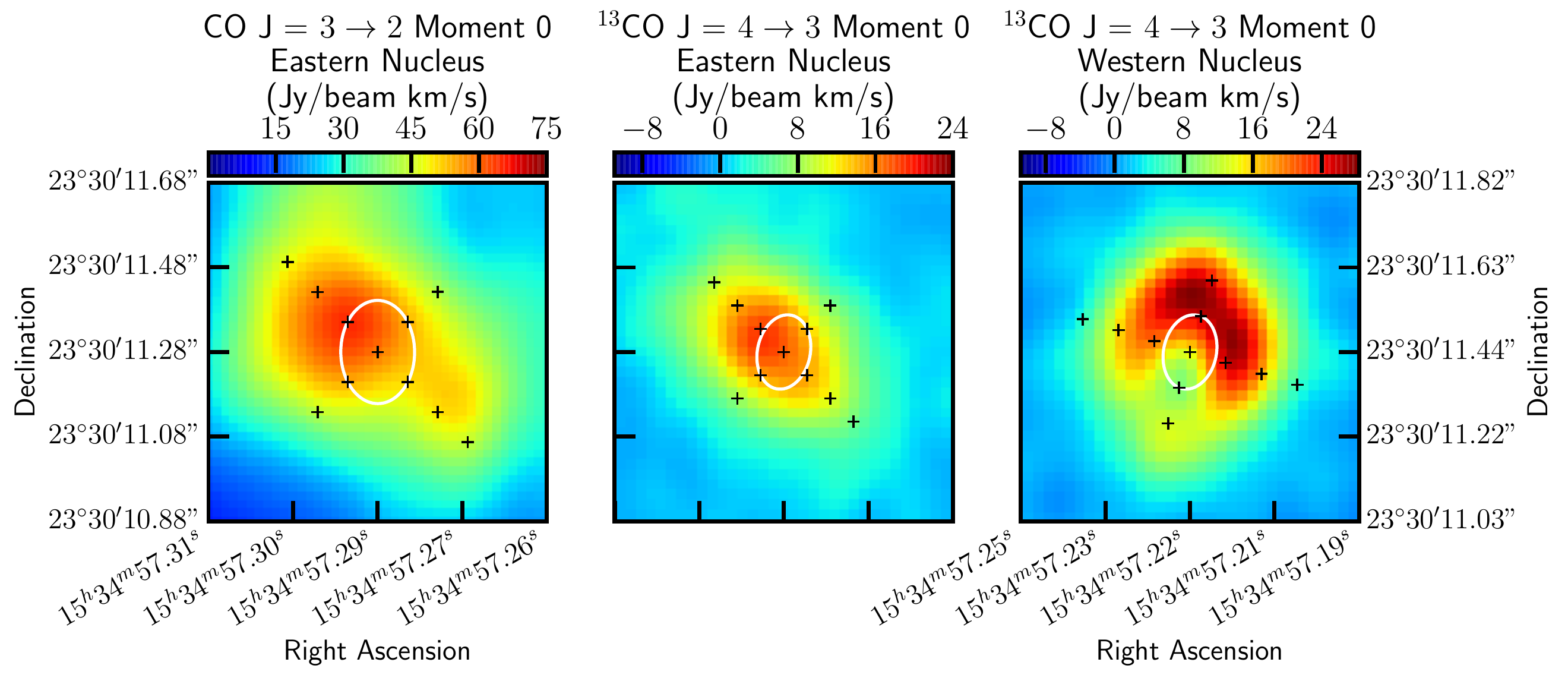}
\caption{Moment 0 for $^{12}$CO J = $3 \rightarrow 2$ eastern nucleus, moment 0 for $^{13}$CO J = $4 \rightarrow 3$ eastern nucleus, and moment 0 for $^{13}$CO J = $4 \rightarrow 3$ western nucleus. Locations where the kinematic profiles were extracted from the data are shown as black crosses. Kinematic profiles were extracted at half-beam increments along the disk's rotational axis and the perpendicular axis. Sampled locations are spaced by the largest of the half-beam increments of the two sampled axes. Beam sizes are shown in white.  \label{fig:sample_locations}}
\end{figure}

After an image is produced by LIME, CASA is used to convolve the image with a Gaussian representation of the observational beam. 
CASA's imcontsub routine is then used to subtract the dust continuum of the image such that the effects of continuum subtraction will be identical to those in the observations.
The spectra are extracted from the model images and the observations at half-beam increments along the plane of the disk as well as orthogonal to it, Figure \ref{fig:sample_locations}.
For both the observations and the model images, the center spectra are taken to be the pixel that is closest to the maximum of the continuum image determined by a Gaussian fit to the continuum emission.  

To produce the disk model for LIME a density profile given by the following equation is assumed.

\begin{equation}
\label{density_eqn}
\rho (r,z) = A \cdot e^{-r/r_{sh}} \cdot e^{-|z|/z_{sh}} \textrm{,}
\end{equation}

where $r^2 = x^2 +y^2$ and A is such that the integral of equation \ref{density_eqn} is the total gas mass, $M_{gas}$. 
The abundance of $^{12}$CO to H$_2$ is assumed to be $2 \times 10^{-4}$. 
The gas temperature, dust temperature, and turbulent velocity are taken to be uniform throughout the disk. 
For the models, the integrated intensity in the continuum map that is produced is required to be the same as that measured in the observations.
In this way, the continuum and gas emissions are simultaneously modeled. 
This constrains the dust temperature for a given gas mass as determined by the model; yet, other dust temperatures could result if different disk geometries and orientations are modeled. 
The default dust opacity values of LIME were used in the models, which are from \citet{Ossenkopf1994}. This gives dust opacities of 2.0 and 2.6 cm$^2$/g for the $^{12}$CO and $^{13}$CO wavebands, respectively.

\subsection{Arp 220 Eastern Nucleus}
The Arp 220 eastern nucleus exhibits asymmetric line profiles with the red side being brighter by a factor of order two for $^{12}$CO.
This is less pronounced in the $^{13}$CO observations, yet, asymmetry is still present.
Several possible causes were previously proposed including a strong outflow or a blueshifted absorber \citep{Rangwala2015}.
In addition to these two explanations, this paper explores another, that the asymmetric profiles are produced by an asymmetric temperature distribution throughout the disk. 
This asymmetric temperature distribution is supported by the very high-resolution images of the eastern nucleus in $^{12}$CO J = $1 \rightarrow 0$ where the non-uniformity of the eastern nucleus is clearly visible \citep{Scoville2017}. 
The spectral profiles of the eastern nucleus's models and observations were sampled along the disk axis taken to be 45$^\circ$, as well as perpendicular to this axis. 

\subsubsection{Foreground Absorber Model}
In order to reduce the flux on the blue side of the eastern nucleus, a blue shifted foreground absorber can be invoked to preferentially absorb emission from the blue side of the disk.
To simulate the absorption, the approach from \citet{Rangwala2015} was used.
A Gaussian absorption profile is invoked with a line center velocity $v_{abs}$, a velocity dispersion $\sigma_{abs} = 90$ km/s, and a line center optical depth $\tau_{abs}$,
which are all allowed to vary with position. 
By tuning the depth and velocity of such an absorber the line profiles seen in the $^{12}$CO J = $3\rightarrow2$ profiles can be easily reproduced. 
However, if this explanation is correct, it must reproduce all of the line profiles for each of the CO lines including the $^{13}$CO line species. 
For the less abundant species of $^{13}$CO, it is expected that the abundance of the absorber should decrease similarly as the abundance ratio of $^{12}$CO to $^{13}$CO in the disk.
If this condition is enforced then less absorption is expected for the $^{13}$CO.
Examining the model results for $^{13}$CO $4 \rightarrow 3$ in Figure \ref{fig:east_models} panel A shows that as the abundance of the absorber is decreased for $^{13}$CO by a factor of 40 while keeping the velocity structure the same, asymmetric line profiles are no longer produced (especially in the green position, Figure \ref{fig:east_models}).
Thus, the asymmetric line profiles observed in $^{13}$CO J = $ 4 \rightarrow 3 $ rule out the possibility of a foreground absorber as the cause of the asymmetry.
For this model, all of the $^{12}$CO J = $3 \rightarrow 2$ gas has a temperature of 110 K and the $^{13}$CO J = $4 \rightarrow 3$ gas has a temperature of 950K.
With the exception of these two temperatures, all of the rest of the parameters are the same as those listed in Table \ref{east_table} for the asymmetric disk model.

\subsubsection{Outflow Model}
Another way to produce asymmetric line profiles is with a strong outflow. 
A strong outflow can skew the line profiles for both the $^{12}$CO species and the $^{13}$CO species as seen in the models in Figure \ref{fig:east_models} panel B.
However, in order to produce such a significant perturbation on the line profiles, an outflow of 150 km/s oriented in the plane of the disk pointed directly at the observer with a 120 degree opening angle was required.  
The requirement that the outflow is oriented in the plane of the disk suggests that this is not a realistic solution.
It does, however, reproduce the central line profiles (black curves in Figure \ref{fig:east_models}) very well for both the $^{12}$CO and the $^{13}$CO species.
It fails, though, to capture the asymmetry in brightness from red to blue when moving further away from the axis of rotation. 
Thus, it is concluded that an outflow cannot be solely responsible for producing the asymmetry from red to blue observed in the line profiles. 
This does not rule out the existence of an outflow.
This is of particular interest when trying to interpret the results from unresolved emission.
If the disk was not resolved and only a single kinematic line profile of a point source was obtained, then an outflow could have reproduced the asymmetric line profiles observed.
This provides a word of caution for interpreting unresolved sources, and for resolved sources demonstrates the usefulness of considering the non-centralized line profiles.
For this model all of the $^{13}$CO J = $4 \rightarrow 3$ gas has a temperature of 1700 K and a total gas mass of 1.4 $\times$ 10$^8$ M$_\odot$.
With the exception of these two parameters, all of the rest of the parameters are the same as those listed in Table \ref{east_table} for the asymmetric disk model. Finally, the fact that such a high gas temperature is required is another suggestion that this model is not an acceptable explanation for the observations.

\subsubsection{Asymmetric Disk Model}
\begin{figure}[ht!]https://www.overleaf.com/project/5c34ecd34ffb762d2081ccd0
\plotone{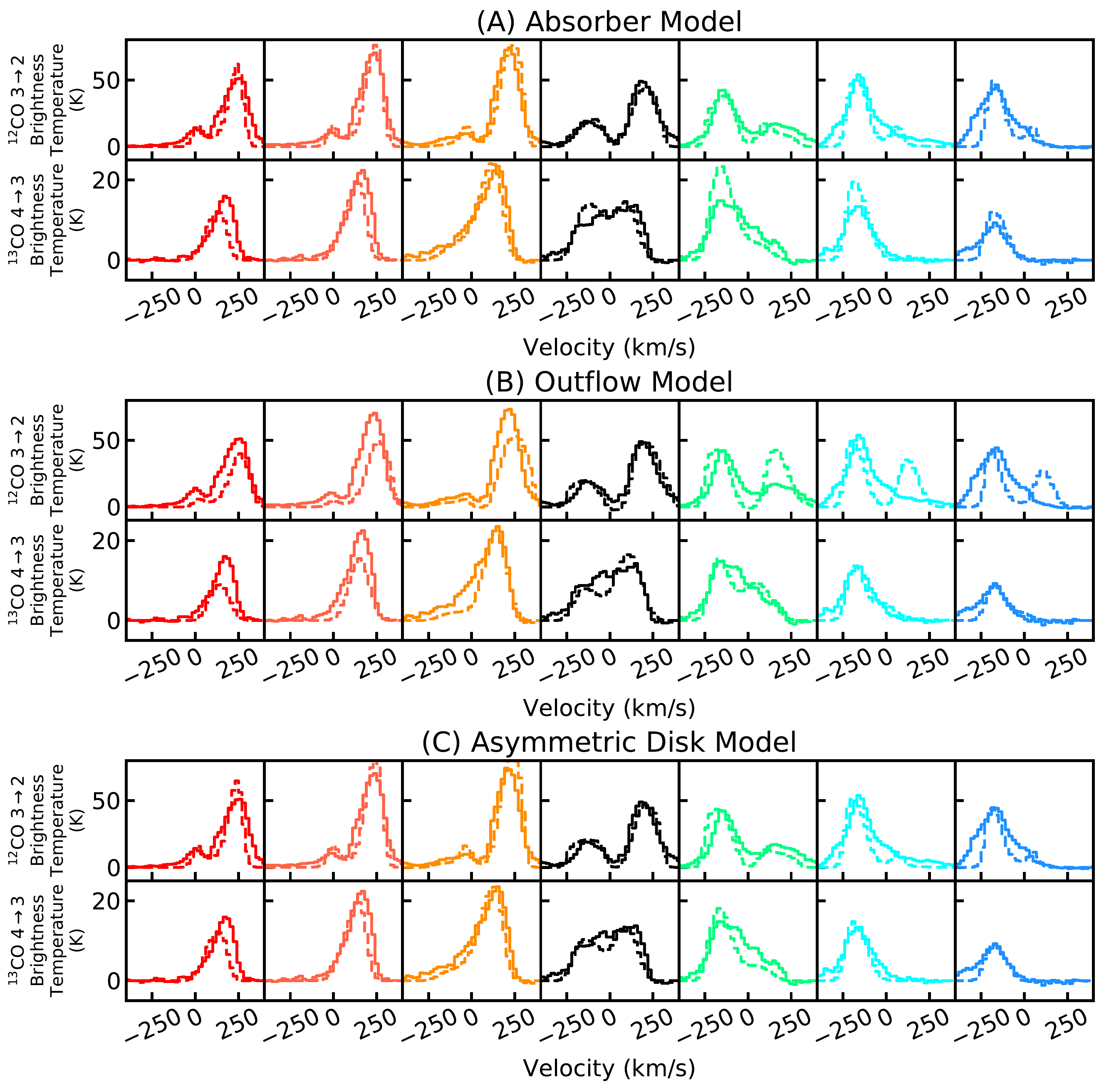}
\caption{Models for Arp 220 east (dashed lines) shown along with the observed line profiles at half-beam increments along the disk axis. $^{12}$CO J = $3 \rightarrow 2$ is shown on top and $^{13}$CO J = $4\rightarrow 3$ is shown on the bottom of each panel.
The column density of the model in the central position is 1.3$\times10^{25}$ H$_2$~cm$^{-2}$ for $^{12}$CO in all of the models.
 The column density of the model in the central position is 3.3$\times10^{24}$ H$_2$~cm$^{-2}$ for $^{13}$CO for both the absorber and asymmetric disk models and in the outflow model is  2.3$\times10^{24}$ H$_2$~cm$^{-2}$.
Moving either to the left or right one panel at a time, the models' column density decreases to 8.3$\times10^{24}$ H$_2$~cm$^{-2}$, 4.2$\times10^{24}$ H$_2$~cm$^{-2}$, and 2.0$\times10^{24}$ H$_2$~cm$^{-2}$ for all of the $^{12}$CO models.
For $^{13}$CO the column density decreases to 2.2$\times10^{24}$ H$_2$~cm$^{-2}$, 1.2$\times10^{24}$ H$_2$~cm$^{-2}$, and 6.5$\times10^{23}$ H$_2$~cm$^{-2}$ for both the absorber and asymmetric disk models.
For the outflow model in $^{13}$CO, the column density decreases to 1.6$\times10^{24}$ H$_2$~cm$^{-2}$, 8.6$\times10^{23}$ H$_2$~cm$^{-2}$, and 4.5$\times10^{23}$ H$_2$~cm$^{-2}$ moving one panel at a time to the left or right.  
For  $^{12}$CO J = $3\rightarrow 2$, 1 Jy/beam = 248 K and for $^{13}$CO J = $4\rightarrow 3$, 1 Jy/beam = 294.5 K. Used in the fitting, but not shown here, are several points perpendicular to the axis of the disk that allows for the constraint of the height of the disk. This is shown in detail for the asymmetric disk model in Figure \ref{fig:mod_east}. \label{fig:east_models}}
\end{figure}

An asymmetric temperature distribution will lead to an asymmetric brightness from one side of the disk to the other.
In this model, for simplicity, the northeastern side of the disk is simply varied to have a temperature that is 60 percent greater than the southwestern side of the disk in $^{12}$CO and 100 percent greater in $^{13}$CO. 
This invokes a sharp discontinuity in the temperature of the model at the center of the disk, but this discontinuity is naturally smoothed by the beam convolution.
It is found that the observed line profiles are best reproduced for this model with an asymmetric temperature distribution, Figure \ref{fig:east_models} panel C. 

Any asymmetries in this nucleus should be smoothed out as the disk rotates.
The fact that the northeastern side of this nucleus is brighter is observationally evident when looking at the $^{12}$ CO $3 \rightarrow 2$ moment 0 map in Figure \ref{fig:12COmom} and even more so in the high resolution $^{12}$ CO $J = 1 \rightarrow 0$ observations from \citet{Scoville2017}. 
Based on the model velocities for this nucleus and the $\sim$ 45 pc radius, the rotation period should be on order $10^5$ years. 
This suggests that the energy source powering this asymmetry must be recent/short-lived. 
This is too fast in comparison to the typical starburst age of $\sim 10^7-10^8$ years for a sustained starburst to be the explanation.
A back-of-the-envelope energy calculation for the required energy to produce a $\Delta T = 45$ K for half of the disk mass, $4 \times 10^8$ M$_\odot$, using K $= \frac{3}{2}N k_B \Delta T$ yields K $=4 \times 10^{51}$ ergs or 4 supernovas worth of energy. 
Likewise, the same calculation for the hotter $^{13}$CO gas yields a bit more energy, $12 \times 10^{51}$ ergs or 12 supernovae.  
It seems reasonable, then, to conclude that this brightness asymmetry could be the result of several recent supernovae.  
Given that the Arp 220 system has a rate of 4 $\pm 2$ supernovae per year \citep{Lonsdale2006}, this seems like a plausible power source. 

\begin{figure}[ht!]
\plotone{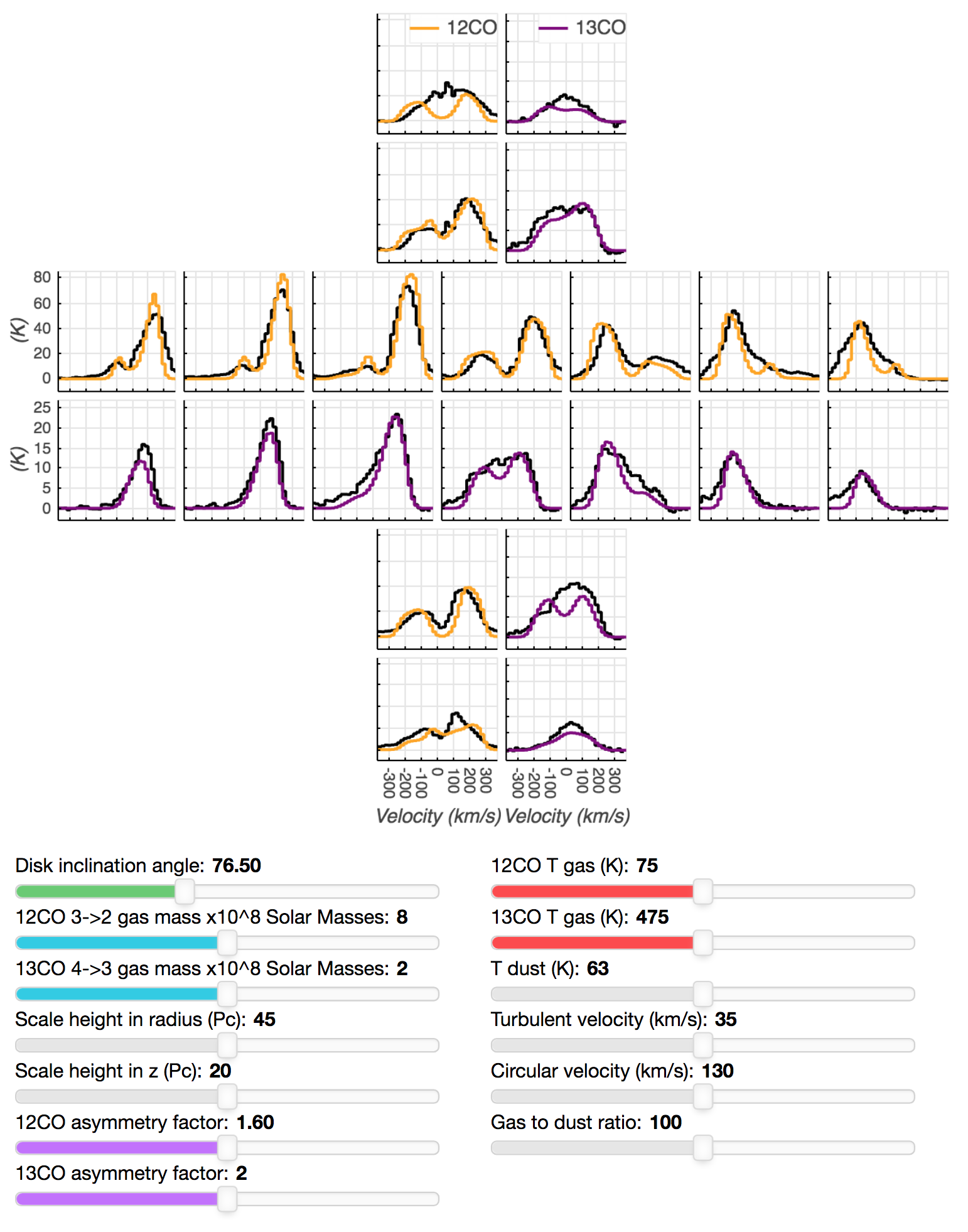}
\caption{Models for the Arp 220 eastern nucleus in $^{12}$CO J = 3-2 (orange lines) and $^{13}$CO J = 4-3 (purple lines) shown along with the observed line profiles at half-beam increments along the disk axis and perpendicular to the disk axis.
The column density of the model in the central position is 1.3$\times10^{25}$ H$_2$~cm$^{-2}$ for $^{12}$CO and 3.3$\times10^{24}$ H$_2$~cm$^{-2}$ for $^{13}$CO. Moving either to the left or right one panel at a time the models' column density decreases to 8.3$\times10^{24}$ H$_2$~cm$^{-2}$, 4.2$\times10^{24}$ H$_2$~cm$^{-2}$, and 2.0$\times10^{24}$ H$_2$cm~$^{-2}$ for $^{12}$CO and to 2.2$\times10^{24}$ H$_2$~cm$^{-2}$, 1.2$\times10^{24}$ H$_2$~cm$^{-2}$, and 6.5$\times10^{23}$ H$_2$~cm$^{-2}$ for $^{13}$CO. Likewise, moving one panel at time in the vertical direction the column density decreases to 2.9$\times10^{24}$ H$_2$~cm$^{-2}$ and 3.1$\times10^{23}$ H$_2$~cm$^{-2}$ for $^{12}$CO and to 9.3$\times10^{23}$ H$_2$~cm$^{-2}$ and 1.4$\times10^{23}$ H$_2$~cm$^{-2}$ for $^{13}$CO. 
For  $^{12}$CO J = $3\rightarrow 2$ 1 Jy/beam = 248 K and for $^{13}$CO J = $4\rightarrow 3$ 1 Jy/beam = 294.5 K. Interactive figure available on-line. \label{fig:mod_east}}
\end{figure}

Considering the preferred fit parameters for this model, shown in Table \ref{east_table}, a few conclusions can be drawn. 
The fitted gas mass of $8 \times 10^8$ M$_\odot$ is close but lower by a factor of two than the results of the derived dynamical mass from \citet{Scoville2017} in CO $J = 1 \rightarrow 0$ of $\sim 1.5 \times 10^9$ M$_\odot$ and \citet{Sakamoto1999} in CO $J = 2 \rightarrow 1$ of $\sim 2 \times 10^9$ M$_\odot$. 
In addition, the column density of the model is calculated via numerical integration of the density formula, Equation \ref{density_eqn}, along the line of sight for each sampled position. 
The values are listed in the caption of Figure \ref{fig:mod_east}.
Central position column densities of 1.3$\times10^{25}$ H$_2$~cm$^{-2}$ for $^{12}$CO and 3.3$\times10^{24}$ H$_2$~cm$^{-2}$ for $^{13}$CO were found.
The $^{12}$CO is in very good agreement with the results of \citet{Gonzalez2012} who also found 1.3$\times10^{25}$ H$_2$cm$^{-2}$, and with \citet{Sakamoto2008} who found $\sim10^{25}$ H$_2$cm$^{-2}$.
The agreement with these other values in the literature help validate this model against other physical measurements of Arp 220 east.

The modeling prefers a thick disk with a vertical  e-folding distance slightly less than a factor of two compared to the radial e-folding distance. 
This is consistent with either a thick turbulent disk or launching of gas to high latitudes in the disk through outflows. 
A turbulent velocity dispersion of $\sigma = $ 43 km/s was found, which is smaller than that found in \citet{Rangwala2015} of 85 km/s. 
It could be that the higher temperature gas probed by $^{12}$CO J = $6 \rightarrow 5$ is more turbulent. 
Alternatively, the large turbulent velocity derived could be the result of \citet{Rangwala2015}'s assumed edge-on geometry which results in narrower line profiles than an inclined disk geometry.
For the preferred fit values the temperatures of the J = $3\rightarrow 2$ gas (probed with $^{12}$CO) and the J = $4 \rightarrow 3$ gas (probed with $^{13}$CO) was found to be dramatically different at 75 K and 475 K.
This is not entirely unexpected because these transitions are on the turning point for which the emission of the entire Arp 220 system goes from cold to warm gas dominated (see Figure 5 from \citet{Rangwala2011}). 
For two such similar transitions, intuitively, the gas is expected to be closer in temperature. 
As such, it is concluded that the hotter temperature in $^{13}$CO may be an effect of the lower optical depth of the $^{13}$CO observations. 
Since $^{13}$CO is optically thinner than $^{12}$CO, it is probing gas deeper in the nucleus where the gas temperatures are potentially hotter.
For the $^{12}$CO, much of the hot gas may be shielded by cooler gas in the foreground.
This effect for the observed dust temperatures given optical depth effects in the continuum emission is discussed in detail in  \citet{Scoville2017,Gonzalez2019}.
Given this, a single temperature gas model may be too simplistic to accurately capture the gas properties in this nucleus.
In addition to this, when adopting an abundance ratio between $^{12}$CO and $^{13}$CO of 40, a lower gas mass for the $^{13}$CO species was also required.
This is consistent with the warm gas component being less massive \citep{Rangwala2011}.
When scaling the gas mass, the dust-to-gas ratio is scaled appropriately so that the total dust mass is kept the same for the two different models.

\begin{longrotatetable}
\begin{deluxetable}{c|cccc}
\tablecaption{Parameters for Arp 220 East Asymetric Disk LIME Model \label{east_table}}
\tablehead{
\colhead{Parameter} & \colhead{Value} & \colhead{Approximate Range}  & \colhead{Description} & \colhead{Defining Spectral Feature(s)}
}
\startdata
$M_{gas}$ & $8 \cdot 10^8$, M$_\odot$ & $4 - 8 \cdot 10^8$  M$_\odot$ & Mass of H$_2$$^a$  & Brightness and depth of  absorption  \\
$M_{gas}$ $^{13}$CO & $2 \cdot 10^8$, M$_\odot$ & $1.6 - 2.8 \cdot 10^8$  M$_\odot$ & Mass of H$_2$  & Brightness and depth of  absorption  \\
$r_{sh}$ & 45 Pc  & 30-52.5 pc & e-folding of density in disk radius & Brightness at large radii\\
$z_{sh}$ & 20 Pc & 15-30 pc & e-folding of density in disk height & Absorption depth \\
$T_{gas}$ & 75, 120 K & 65-85, 104-136 K & Gas temperature$^{a}$ & Brightness \\
$T_{gas}$ $^{13}$CO & 475, 950 K & 375-575, 750-1150 K & Gas temperature$^{a}$ & Brightness \\
$T_{dust}$ & 63 & 50-80$^b$ & Dust temperature$^{c}$ & Absorption depth, continuum brightness\\
$^{12}$CO/$^{13}$CO  & 40 & Not well constrained$^{d}$ &  Abundance of $^{12}$CO to $^{13}$CO$^a$ & Brightness\\
$v_{turb}$ & 43 km/s & 38-48 km/s & Turbulent velocity & Line width and doubly peaked lines\\
$v_{circ}$ & 180, 130$^e$  & 160 - 180, 115 - 135 km/s& Circular velocity & Line width and singly peaked lines\\
gIId & 100 & Not constrained & Gas to dust mass ratio M$(H_2)$/M(dust) & Absorption depth\\
$\phi$ & $76.5^\circ$ & 72-90$^\circ$  & Disk inclination angle from face-on & Importance of $v_{turb}$ vrs $v_{circ}$, aspect ratio \\
\enddata
\tablenotetext{a}{All of these parameters are, to some extent, degenerate with each other.}
\tablenotetext{b}{Different dust temperatures require different M$_{gas}$ to correctly fit the total continuum luminosity and thus, the constraints are not reflected by varying this parameter in the interactive figure.}
\tablenotetext{c}{Determined by integrated continuum emission.}
\tablenotetext{d}{Because $^{13}$CO and $^{12}$CO are different species, abundance variations and temperature variations cannot be distinguished from each other.}
\tablenotetext{e}{First value is constant out to 40
pc and decreases linearly to the second value at 150 pc.}

\tablecomments{The approximate range was found by varying only that parameter while all other parameters were kept the same and then by eye deciding an acceptable range of believability. A more quantitative treatment was computationally too expensive. As such, the reader is invited to use the interactive version of Figure \ref{fig:mod_east} (available on-line) to come to his or her own conclusions about which range is believable for a good fit. For parameters that are degenerate, a few multidimensional models are shown in the appendix so that the degeneracies can be explored.}
\end{deluxetable}
\end{longrotatetable}

\subsection{Arp 220 Western Nucleus}
The western nucleus in $^{12}$CO J = $3 \rightarrow 2$ exhibits 3 separate peaks in the spectrum and an extremely narrow absorption feature, Figure \ref{fig:lines}.
The narrow absorption dip is believed to be the result of a small-scale velocity-coherent gas complex in the outer envelope of the western nucleus gas. 
The spectral features observed in $^{12}$CO J = $3 \rightarrow 2$ cannot be reproduced by the smooth disk model approximation used in this paper. 
As such, modeling the western nucleus in $^{12}$CO J = $3 \rightarrow 2$ is not attempted.
In $^{13}$CO J = $4 \rightarrow 3$ the spectra become smoother, exhibiting only two separate peaks that would be expected for a disk. 
Thus, it is concluded that the $^{13}$CO J = $4 \rightarrow 3$ is probing the nuclear disk as a whole better than the $^{12}$CO J = $3 \rightarrow 2$ and that the resultant spectra are compatible with the smooth disk model used in this paper. 
This does not mean that the $^{13}$CO J = $4 \rightarrow 3$ observations are probing the nucleus in its entirety.
Rather, the $^{13}$CO line profile (Figure \ref{fig:lines}) exhibits a large central absorption dip, indicating a high optical depth that is absorbing continuum radiation from the nucleus core.

The spectral profiles of the western nucleus were sampled at 3 half-beam distance increments from the center of the continuum along an axis of 17$^\circ$, Figure \ref{fig:sample_locations}.
17$^\circ$ is the deconvolved axis derived from the high spatial resolution observations from \citet{Scoville2017}. 
In addition, the spectral profiles are sampled perpendicular to this axis for two half-beam increments. 
For the western nucleus, it was found that a constant circular velocity, $v_{circ}$, captured the velocity structure of the disk well enough that a velocity that varies with radius was not required.
A prior was taken that the disk inclination angle is such that the red outflow will be directed to the northwest and the blue outflow toward the southeast, supported by the observed outflow in $^{12}$CO J $ =3 \rightarrow 2$ and the outflow seen in HCN \citep{Barcos2018}.
The outflow velocity modifies the systematic rotational velocity in the disk within the cone defined by the opening angle, $\theta_{out}$ such that the velocity within this cone is a combination of the rotational velocity as well as the outflow velocity.
However, for the blue outflow, in order to obtain a narrower absorption feature within the outflow cone, $v_{out}$ is assumed to be the only systematic velocity without any rotational velocity. 

\begin{figure}[ht!]
\plotone{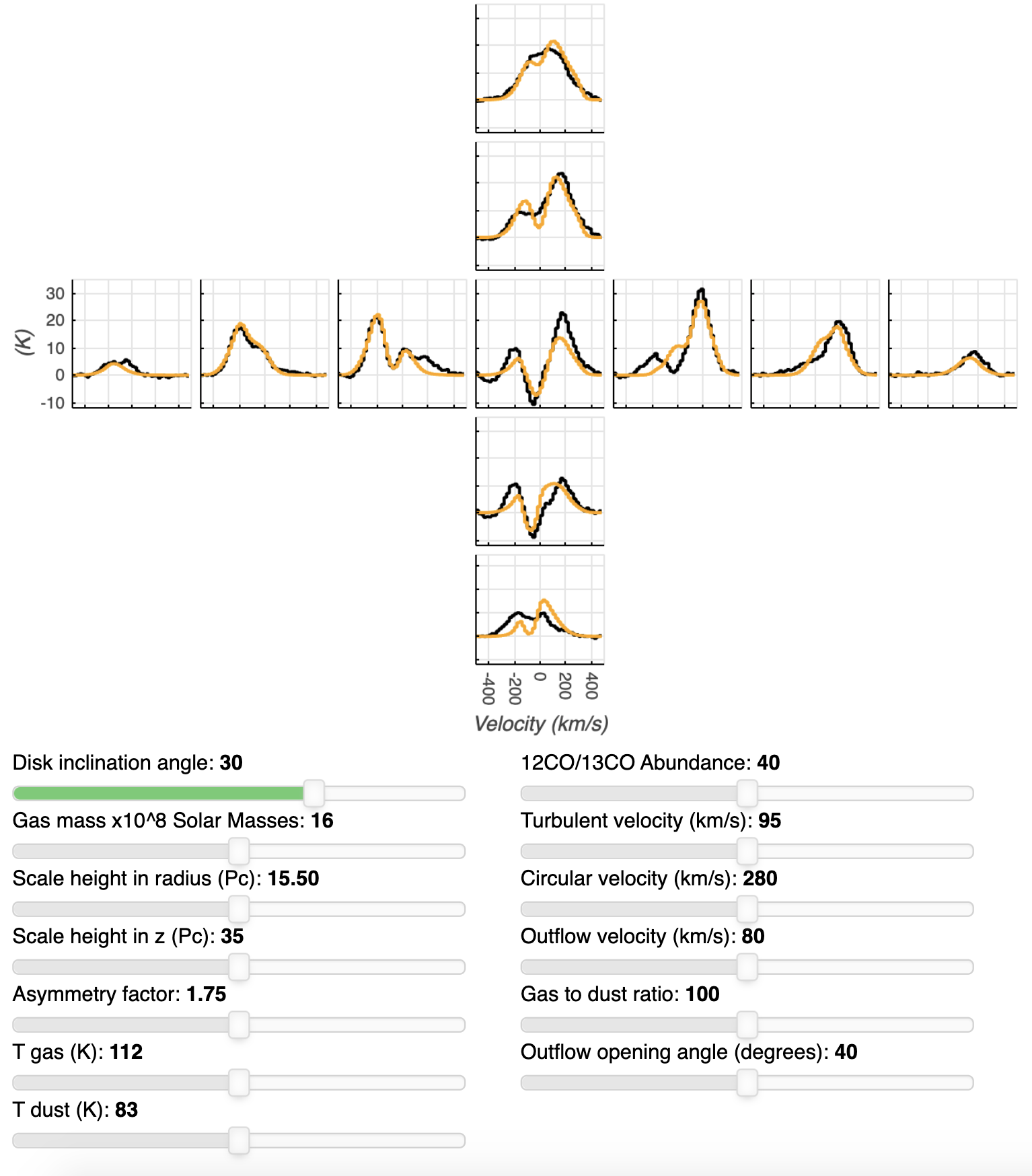}
\caption{Models for the Arp 220 western nucleus in $^{13}$CO J = 4-3 (orange lines) shown along with the observed line profiles at half-beam increments along the disk axis and perpendicular to the disk axis.  The column density of the model in the central position is 4.9$\times10^{25}$ H$_2$~cm$^{-2}$. Moving either to the left or right one panel at a time the models' column density decreases to 1.1$\times10^{25}$ H$_2$~cm$^{-2}$, 1.5$\times10^{24}$ H$_2$~cm$^{-2}$, and 2.0$\times10^{23}$ H$_2$~cm$^{-2}$. Likewise, moving one panel at time in the vertical direction the column density decreases to 9.5$\times10^{24}$ H$_2$~cm$^{-2}$ and 1.1$\times10^{24}$ H$_2$~cm$^{-2}$.  1 Jy/beam = 294.5 K. Interactive figure available on-line. \label{fig:west}}
\end{figure}

The western nucleus is reasonably well modeled by a simple disk with both a red and blue outflow and a few asymmetries. 
The observations and the model are shown in Figure \ref{fig:west} and the derived parameters are shown in Table \ref{west_table}.
In addition, the column density of the model is calculated via numerical integration of the density formula, Equation \ref{density_eqn}, along the line of sight for each sampled position. 
The values are listed in the caption of Figure \ref{fig:west}.
%Jason wants a description of how the model fits 
A few notable conclusions from the fitted parameters are drawn below:\\
1. The fitted gas mass of $1.6 \times 10^9$ M$_\odot$ agrees closely with the results of the derived dynamical mass from \citet{Scoville2017} in CO $J = 1 \rightarrow 0$ of $\sim 1.5 \times 10^9$ M$_\odot$ and \citet{Sakamoto1999} in CO $J = 2 \rightarrow 1$ of $\sim 2 \times 10^9$ M$_\odot$. 
This is somewhat surprising because for the $J = 4 \rightarrow 3$ transition one might expect based on the results of single dish experiments \citep{Rangwala2011} that these observations should be probing a less massive warm gas component.
However, as discussed for the eastern nucleus, this may be an effect of the optical depth of the observations.
Since the optical depth is large, the $^{13}$CO J = $4 \rightarrow 3$ transition may be just probing a cooler outer envelope of gas in the nucleus.
Ultimately, transitions that trace cool, warm, and the combination of cool and warm gas need to be modeled simultaneously for each nucleus to completely understand the significance of cool versus warm gas for different transitions.
The derived gas masses are dependent on the $^{12}$CO/$^{13}$CO ratio, in this case, set to 40, which is a number that cannot be robustly constrained with these observations. 
It is worth noting that if the observations are probing a lower mass, hotter component of the gas, then the gas-to-dust ratio in the model should be adjusted to be less than 100 in order to reflect the true dust to total gas mass ratio.
If this is the case,  then the derived dust temperature would be lower than that reported in Table \ref{west_table}. 
However, attempts to produce models with a lower gas-to-dust ratio failed to reproduce the correct continuum flux density while maintaining the strong central absorption feature.
Additionally to the modeled mass, we find a central position column density of 4.9$\times10^{25}$ H$_2$~cm$^{-2}$.
This is in good agreement with the results of \citet{Gonzalez2012}, who found 3-6$\times10^{25}$ H$_2$~cm$^{-2}$, and is in approximate agreement with \citet{Sakamoto2008} who found $\sim10^{25}$ H$_2$~cm$^{-2}$.
This is a factor of 4 less than the 2$\times10^{26}$ H$_2$~cm$^{-2}$ found by \citet{Scoville2017}. However, that column density measurement is derived from the dust measurements whereas these are derived from modeling the CO gas, both of which require their own different assumptions which can effect the derived values. 
The rough agreement with the other values in the literature do help validate this model against other physical measurements of Arp 220 west. \\
2. $z_{sh}>r_{sh}$: Fitting the western nucleus requires gas at high latitudes in the disk. 
This is most strongly constrained by the depth of the central absorption.
The western nucleus, with an inclination angle of 30$^\circ$, is much closer to face-on than the eastern nucleus.
The result of this is that the observations do not probe the $z_{sh}$ spatially, which degrades the ability to constrain it with the models. 
However, the parameter $z_{sh}$ is indirectly probed by the depth of the absorption feature.
Through the models, it is found that recreating the central absorption dip is best facilitated by having gas at high latitudes in the disk, a larger $z_{sh}$. 
However, as discussed in the appendix, when trying to recreate a central absorption dip, $z_{sh}$ is degenerate with the ratio of the gas to dust temperatures, the total gas mass, and the inclination angle. 
This further degrades the accuracy of the derived $z_{sh}$.
As such, it is possible if the model were modified to have two temperature components with a lower temperature further out from the nuclear core, a smaller $z_{sh}$ might be obtained.
With only one transition, invoking a second temperature in the model would overfit the observations without another transition that traces a different temperature component of the gas to constrain a two-temperature model.
This does suggest that more transitions would potentially help achieve better constraints for $z_{sh}$.
$z_{sh}>r_{sh}$ invokes an unrealistic aspect ratio, and as such, it can be concluded that the models do not accurately constrain the aspect ratio of the disk.
However, the models do indicate that gas at high latitude in the disk is important for recreating the central absorption dip observed in the observations.
This potentially indicates the presence of either foreground gas or that outflows are injecting gas into very high latitudes.\\
3. The narrow absorption dips require a smaller turbulent velocity gas component (compared to the disk gas) in the foreground to be accurately reproduced.
This further indicates that the high latitude gas has different characteristics than the rest of the disk gas. 
Either the disk has a parameter gradient with latitude or the central absorption dip traces a gas complex that is distinct from the disk gas.
Furthermore, the strong central absorption dip demonstrates that even the $^{13}$CO observations are probing gas that is optically thick and absorbing. 
This leads to the next conclusion. \\
4. $^{13}$CO J = $4\rightarrow 3$ is optically thick. 
Using LIME, an optical depth map for the model shown in Figure \ref{fig:west} was produced.
In this map, a peak optical depth at the center of the western nucleus is found to be 18 and if the dust is removed from the model a $\tau$ of 15 is obtained. 
Furthermore, changing model parameters within 20 pc of the center of the disk has no effect on the model's produced line profiles, demonstrating an inability to probe to the very center of the disk with these observations. 
Thus, the absorption is obscuring information from deep inside the western nucleus.
In this situation, the modeling would greatly benefit from an even less abundant species such as C$^{18}$O.
However, observations of less abundant tracers will eventually be limited by the optical depth of the dust. \\
5. The line profiles at the center of the western nucleus dip below zero.
Thus, it must be concluded that the $^{13}$CO is, in fact, absorbing the dust emission. 
There is likely also a small amount of CO self-absorption occurring; however, when the dust is removed in the model only a small central dip is observed, suggesting minimal CO self-absorption.
These models do suggest that this is not the case for the $^{12}$CO species where CO self-absorption does become significant in the western nucleus. \\
6. The modeled outflows help predict the blue-shifted absorption in the center of the western nucleus and the high-velocity red excess in the northwest of the western nucleus.
The outflows derived with this model are more modest than the outflow that is detected in the high-velocity emission in $^{12}$CO $J = 3 \rightarrow 2$ observations. 
It is believed these are all part of the same outflow but the $^{13}$CO observations do not have the sensitivity and bandwidth to detect the higher velocity components of the outflow.   

\begin{longrotatetable}
\begin{deluxetable}{c|cccc}
\tablecaption{Parameters for Arp 220 West LIME Model \label{west_table}}
\tablehead{
\colhead{Parameter} & \colhead{Value} & \colhead{Approximate Range}  & \colhead{Description} & \colhead{Defining Spectral Feature(s)}
}
\startdata
$M_{gas}$ & $1.6 \cdot 10^9$  M$_\odot$ & $0.8 - 2.4 \cdot 10^9$  M$_\odot$ & Mass of H$_2$$^a$  & Brightness and depth of  absorption  \\
$r_{sh}$ & 15.5 Pc  & 12.5-18.5 Pc & e-folding of density in disk radius$^b$ & Brightness at large radii\\
$z_{sh}$ & 35 Pc & 27.5-50 Pc & e-folding of density in disk height & Absorption depth \\
$T_{gas}$ & 112 K & 120-210 K & Gas temperature$^{a,c}$ & Brightness \\
$T_{dust}$ & 83 & 76-86 & Dust temperature$^d$ & Brightness and depth of absorption\\
$^{12}$CO/$^{13}$CO   & 40 & Not constrained &  Abundance of $^{12}$CO to $^{13}$CO$^a$ & Brightness\\
$v_{turb}$ & 95 km/s & 85-105 km/s & Turbulent velocity$^e$ & Line width and doubly peaked lines\\
$v_{circ}$ & 280 km/s & 250-310 km/s& Circular velocity & Line width and singly peaked lines\\
$v_{out-red}$ & 150 km/s & 75-220 km/s & Outflow velocity & High velocity emission\\
$v_{out-blue}$ & 80 km/s & 40-120 km/s & Outflow velocity & High velocity absorption\\
$\theta_{out}$ & 40$^\circ$ & 30-50$^\circ$ & Outflow opening angle & Intensity of high velocity emission\\
gIId & 100 & Not constrained & Gas to dust mass ratio M$(H_2)$/M(dust) & Absorption depth\\
$\phi$ & 30$^\circ$ & 25-35$^\circ$  & Disk inclination angle from face-on & Importance of $v_{turb}$ vrs $v_{circ}$, aspect ratio \\
\enddata
\tablenotetext{a}{Parameters somewhat degenerate with each other.}
\tablenotetext{b}{Value decreases to 12.5 PC for the southeast corner.}
\tablenotetext{c}{Value increases to 196 K for $y > 0$ (north side in image).}
\tablenotetext{d}{Value increases to 162 K at $r < 25$ pc.}
\tablenotetext{e}{Value decreases to 55 km/s at  $|z| > 20$ pc.}
\tablecomments{The approximate range was found by varying only that parameter while all other parameters were kept the same and by eye deciding an acceptable range of believability. A more quantitative treatment was computationally too expensive. As such, the reader is invited to use the interactive version of Figure \ref{fig:west} (available on-line) to come to his or her own conclusions about which range is believable for a good fit. For parameters that are degenerate, a few multidimensional models are shown in the appendix so that the degeneracies can be explored.}
\end{deluxetable}
\end{longrotatetable}

%%%%%%%%%%%%%%%%%%%%%%%%%%%%%%%%%%%%%%%%%%%%%%%%%%%%%%%%%%%%%%%%%%%
%                   Discussion
%%%%%%%%%%%%%%%%%%%%%%%%%%%%%%%%%%%%%%%%%%%%%%%%%%%%%%%%%%%%%%%%%%%%
\section{Discussion}

\begin{figure}[ht!]
\plotone{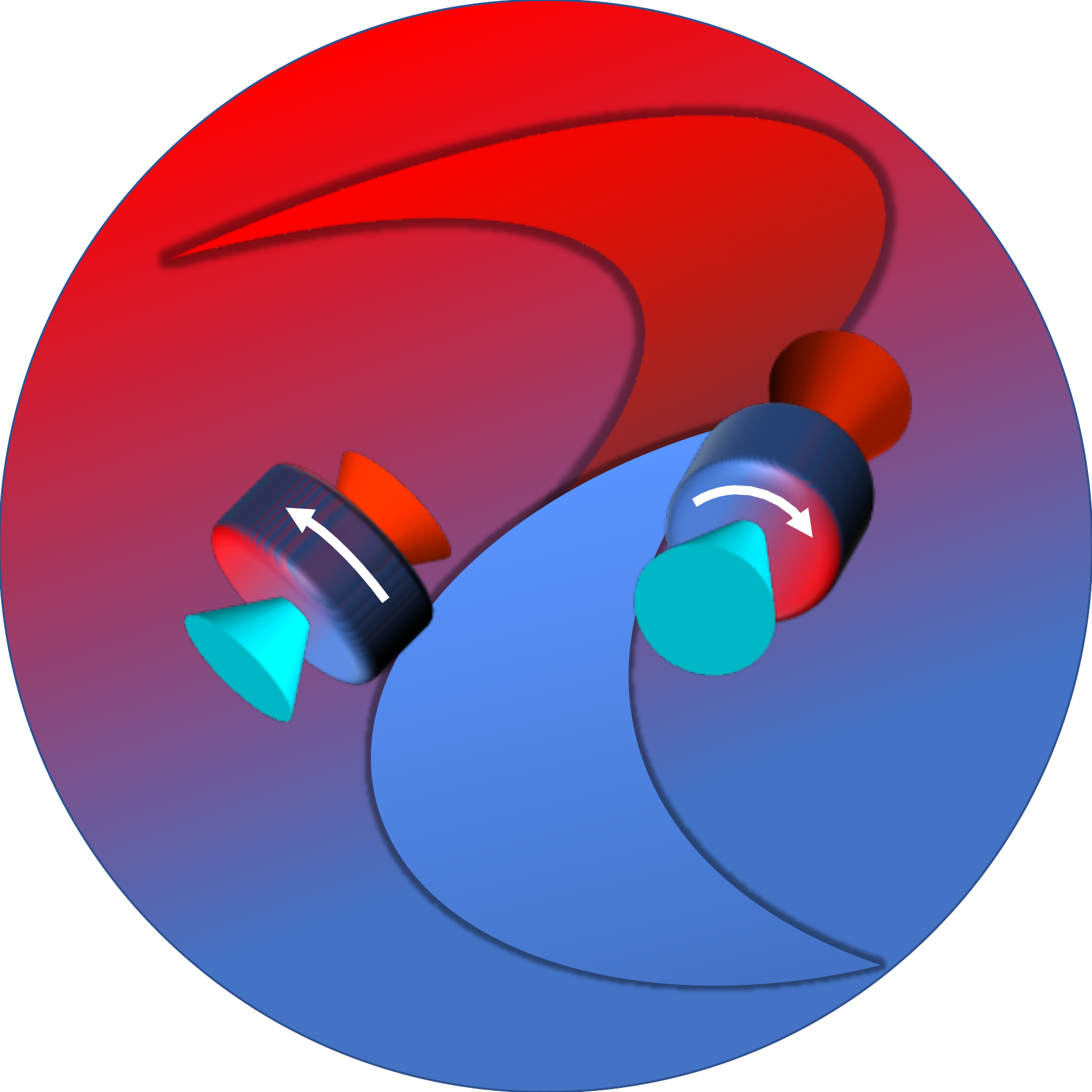}
\caption{ Schematic of Arp 220. 
Thick disks are shown with their inclination angles and outflows indicated.
Redshifted and blueshifted gas are shown with red and blue colors.
The outer extended emission disk of which the inclination angle is not determined is shown with tidal over density features.  \label{fig:cartoon}}
\end{figure}

The results of this analysis reaffirm and extend the results of previous work on Arp 220. 
The general picture of two counter-rotating disks embedded in an extended emission disk is the same as in \citet{Sakamoto1999}. 
The nuclei masses roughly agree with the dynamical masses from \citet{Scoville2017,Sakamoto1999}.
Thick turbulent disks as seen in \citet{Rangwala2011} are found.
The strongly collimated outflow in the western nucleus observed by \citet{Barcos2018} was confirmed in these observations but an additional outflow in the eastern nucleus was also found. 
The 3-D modeling allows for a better understanding of the contributions of turbulent velocity, rotational velocity, and disk inclination angle to the observed kinematic line profiles. 
Though these parameters are somewhat degenerate, these estimates are believed to be accurate to within 20\%, although no statistical fitting was carried out. 
These parameter derivations, along with the observed outflows, allow for the construction of a schematic depicting the rotating disks' orientations, Figure \ref{fig:cartoon}.
In addition, structure is found within the extended disk detected by \citet{Sakamoto1999}.
This structure resembles tidal features suggesting potential inflow of gas to the most central regions of Arp 220. 
An outstanding question is, does such structure exist in other merging ULIRG systems or is this unique to the Arp 220 system? 
It is possible that this inflow is required to provide sufficient gas for star formation to be sustained over observable times scales. 

% Arp 220 as a template for high redshift sources
Arp 220 has often been used as an analog to understand high-redshift galaxies.
Is Arp 220 a good prototype for high-redshift galaxies?
Submillimeter galaxies (SMGs) are bright high-redshift galaxies that have continuum detections of $ \geq$ 1 mJy detected in the range from 250 $\mu$m to 2 mm \citep{Casey2014}. 
Historically, SMGs refer to SCUBA detections at 850 $\mu$m with flux densities of at least 5 mJy \citep{Hughes1998,Barger1998}.
Due to their high luminosity, $\sim 10^{12.5}$ L$_\odot$, astronomers were led to the natural conclusion that these populations represented the equivalent local ULIRGs but at high redshifts. 
This is supported by observational evidence that SMGs often exhibit disturbed morphologies and kinematics (e.g., \cite{Engel2010,Menendez2013}) and that local ULIRGs are the product of enhanced star formation from major merger events. 
At the same time, resolved observations found that many SMGs exhibited morphologies that are much larger than local ULIRGs, which are in general smaller than 1 kpc.
This led to the idea that SMGs may not be merging starburst systems similar to local universe ULIRGs. 
Rather, these galaxies represent the high-mass end of a normal disk galaxy main sequence that has increased star formation efficiency at high-redshift. 
This is in stark contrast to Arp 220 which is clearly a merger driven starburst system that cannot be long-lived given the derived gas depletion times, $\sim$10 Myr. 
A $\sim$10 Myr timescale is also not compatible with the observed number counts of SMGs at high redshift \citep{Dave2010}.
Recent simulations and observations support the conclusion that most SMGs are massive normal disk galaxies rather than Arp 220-like mergers \citep{Narayanan2015,Koprowski2016,Michalowski2017,Scoville2017b}.
However, it does seem that $\sim$5\% of these galaxies have elevated star formation rates that cause them to lie off the normal galaxy main sequence \citep{Sargent2012}. 
These systems can be considered high-redshift starburst galaxies. 
The question that remains now is, are these high-redshift starburst galaxies driven by major mergers like local ULIRG starburst galaxies such as Arp 220 or is their heightened star formation rate driven by increased gas accretion efficiency or minor mergers?

Given this, it is worth considering, using knowledge of local starbursts such as Arp 220, what observational characteristics should be searched for in these high-redshift starbursting galaxies that might indicate if they are the product of mergers or not. 
A natural choice might be to look at the physical size of these galaxies since this is one of the first observational characteristics that suggested SMGs may not be analogs of local ULIRGs.
However, the relatively larger extended star formation is in fact predicted for some merger simulations for the high gas masses present in these high-redshift systems \citep{Narayanan2009}.
The results from this paper present an interesting introspective into this idea. 
In the $^{12}$CO J = $3 \rightarrow 2$ observations, significant extended emission over 1.3 kpc that comprises 50\% of the total emission is detected.
This is somewhat larger than the typical assumed emission scales for local ULIRGs of $<$ 1 kpc, though still relatively compact.
It seems reasonable to suspect that with an increasing gas fraction at high redshift this extended component could be even larger. 
On the other hand, it is worth considering whether these observations completely detect the extended size of Arp 220, or would additional extended gas be revealed by more sensitive lower resolution observations?   
These considerations caution careful use of physical size as an indicator of a merging or non-merging system. 
Ultimately, larger surveys of the most luminous SMGs coupled with resolved spatial kinematic follow-up observations of the gas in these galaxies will unravel whether these systems are mergers or disk galaxies with enhanced gas accretion.
Certainly, if sub-kpc resolution with high S/N can be obtained to individually resolve two different nuclear disks within one SMG, a merger would be unequivocally determined. 
In addition, the tidal features seen here in Arp 220, though probably difficult to detect, would present another possible identification of a merging system.
Detailed understanding of the local ULIRGs, like Arp 220, will aid the community in making this distinction.

\begin{figure}[ht!]
\plotone{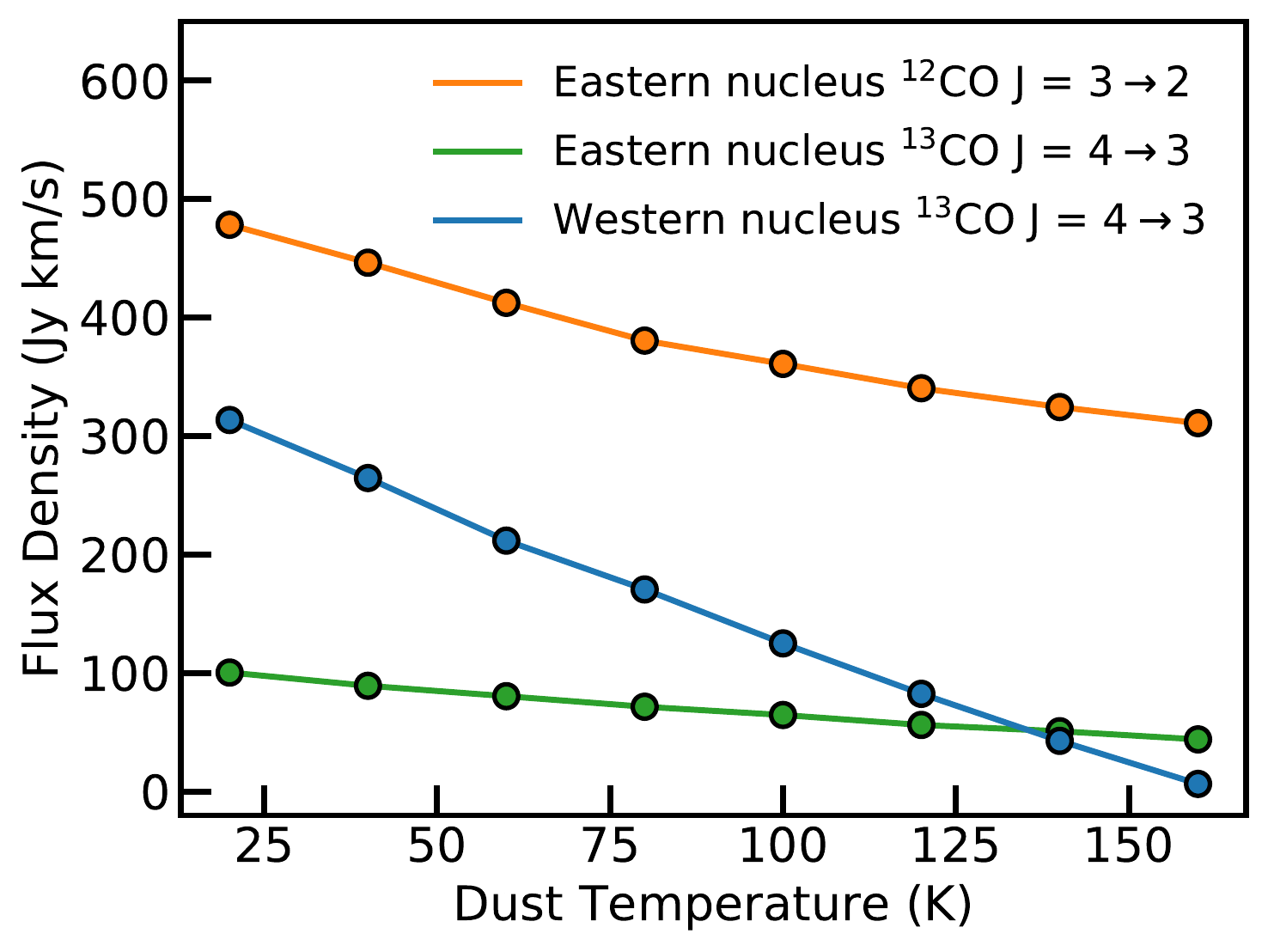}
\caption{Plot showing the dependence of the cumulative flux density of the asymmetric disk model for the eastern nucleus and the model for the western nucleus if the dust temperature is changed. Due to significant continuum absorption in the western nucleus, there is a strong dependence on the CO brightness with dust temperature. A factor of two difference in total flux density is observed as the dust temperature is varied from 20 to 80 K. For the eastern nucleus, this dependence is weaker and is even weaker for the $^{13}$CO J = $4 \rightarrow 3$ line, which is not as optically thick.  \label{fig:flux_v_dust_T}}
\end{figure}

One reason to determine which SMGs are starburst systems driven by mergers, versus normal galaxies, is to obtain accurate gas mass estimates from CO observations. 
The general practice for determining H$_2$ gas mass from CO line luminosity is to use a conversion factor of 0.8 M$_\odot$ (K km/s pc$^2$)$^{-1}$ for starburst systems and to use a Milky Way conversion factor of 4.0 M$_\odot$ (K km/s pc$^2$)$^{-1}$ for normal galaxies \citep{Bolatto2013}. 
While this conversion factor is not measured in this paper for Arp 220, nor is a conversion factor suggested for SMGs, it is observed from the modeling that physical characteristics of the nuclei can significantly impact the CO line luminosity. 
The CO line luminosity may, in turn, impact what the CO to H$_2$ conversion factor value should be. 
In particular, the observations and modeling show that there can be a significant effect on the line luminosity from the extreme optical depth in the Arp 220 nuclei absorbing continuum radiation. 
The absorption of the continuum radiation reduces the observed brightness of the CO emission in Arp 220.
To be clear, the CO is still emitting at zero velocity, but due to the larger absorption of the continuum emission compared to the CO emission, it appears in the continuum subtracted spectra that no CO emission is present. 
This reduced apparent emission is dependent on the dust emission and thus the dust temperature.
This effect is demonstrated in Figure \ref{fig:flux_v_dust_T}. 
If continuum emission absorption was not taken into account, it might be concluded that there is less gas in the Arp 220 nucleus emitting, leading to underestimates of the gas mass.
However, this effect may be canceled out by another. 
For instance, a higher dust temperature leading to reduced CO emission may also always be accompanied by a higher gas temperature leading to more CO emission. 
The results do show that consideration of continuum absorption in line measurements is warranted for systems with large optical depths when deriving line luminosities. 

%%%%%%%%%%%%%%%%%%%%%%%%%%%%%%%%%%%%%%%%%%%%%%%%%%%%%%%%%%%%%%%%%%%
%                   Conclusions
%%%%%%%%%%%%%%%%%%%%%%%%%%%%%%%%%%%%%%%%%%%%%%%%%%%%%%%%%%%%%%%%%%%%

\section{Conclusions}

% wow our observations are great
The observations of Arp 220 in $^{12}$CO J = $3 \rightarrow 2$, $^{13}$CO J = $4 \rightarrow 3$, HCN J = $5 \rightarrow 4$, and SiO J = $8 \rightarrow 7$ reveal incredible amounts of information about the morphology, kinematics, and internal physics of the Arp 220 nuclei and surrounding material.
Extremely bright HCN emission is detected that is spatially different from the CO emission. 
Extremely deep SiO absorption is detected in the western nucleus but only just detected in the eastern nucleus, suggesting shocks are significant in the western nucleus but less so for the eastern. 

% extended emission conclusions
The extended emission detected in $^{12}$CO J $= 3 \rightarrow 2$ is significant in both flux, with an integrated emission greater than both nuclei combined, but also in structure. 
This points to the need for both high-resolution imaging to resolve the nuclear disks and also that lower resolution high sensitivity observations capturing the extended emission are equally important.
The extended emission exhibits a m = 2 asymmetry which may be indicative of tidal torques that are the signature for gas inflow providing the fuel source for the strong nuclear star formation observed. 

%outlfow conclusion
Collimated outflows are again observed for the western nucleus in $^{12}$CO J = $3 \rightarrow 2$ and for the first time are also strongly detected in the eastern nucleus.
This demonstrates the power of these ALMA observations where outflows can be both kinematically and spatially resolved from the nuclear disks.
Both nucleis' outflows are determined to contain significant mass.
These outflows indicate that either star formation or AGN  activity is contributing significant feedback.
%, enriching the interstellar medium of Arp 220 with substantial amounts of gas. 
In addition, the depletion time scales of these outflows suggest that the massive amount of star formation in the Arp 220 system only represents a short time scale event in Arp 220's evolution.  

%modeling conclusion
The kinematic models of the eastern nucleus suggest a highly asymmetric disk in temperature as the best explanation for the asymmetric line profiles observed. 
The inclusion of the less abundant $^{13}$CO species in the models allows for ruling out the possibility of a foreground absorber as the cause of the asymmetric line profiles.
The asymmetry suggests fast time scale energetic events are present in the eastern nucleus, with supernova as a plausible power source. 
In fact, given the frequency of supernova in the Arp 220 system, it seems remarkable that the nuclear disks are not more perturbed.
It was also found that the line profiles of the eastern nucleus cannot be reproduced simply by adding an outflow. 
Asymmetric line profiles may not always indicate the presence of an outflow.  

The modeling of the western nucleus suggests the presence of high latitude absorbing gas in the disk. 
Furthermore, it demonstrates that the disk is highly optically thick, where even the $^{13}$CO species is absorbing significant amounts of continuum radiation.
In addition, modeling in both nuclei suggests that the inferred gas temperatures  may have a dependence on the optical depth of the observed species.
The extreme optical depth of Arp 220 demonstrates that the optical depth, the depth to which observations probe within a gas complex, must be considered when deriving the physical characteristics of the gas such as temperature.
So much so that observations of different abundance tracers of the same transition will not necessarily be probing the same gas. 
Additionally, continuum absorption by the molecular gas lines will reduce the total apparent emission from these lines. 

Finally, it is concluded that a better understanding of the gas kinematics and physical characteristics of the nuclear disks could be obtained by incorporating a less abundant species such as C$^{18}$O to probe deeper into the nuclei. 
In addition, simultaneous modeling of multiple CO transitions that trace  cold gas,  warm gas, and a combination of both the cold and warm gas is needed to develop the best understanding of the gas temperatures and their physical distributions within each nucleus.

This paper makes use of the following ALMA data: ADS/JAO.ALMA\#2015.1.00736.S. ALMA is a partnership of ESO (representing its member states), NSF (USA) and NINS (Japan), together with NRC (Canada), MOST and ASIAA (Taiwan), and KASI (Republic of Korea), in cooperation with the Republic of Chile. The Joint ALMA Observatory is operated by ESO, AUI/NRAO and NAOJ.

\appendix

\section{LIME parameter dependencies}
LIME is efficient but not without computational cost.
This makes producing Monte Carlo simulations to determine error bars for the derived best-fit quantities difficult without the aid of a supercomputer.
However, with a few targeted simulations scaling some of the input variables, one can build up an understanding of how changing a particular variable in LIME effects the output line profiles.
This helps build intuition for how to interpret observed line profiles and gives insights into the confidence for the derived parameters.

Here some of the parameters for the western nucleus are varied to demonstrate how they impact the line profiles.
This is a hands-on experiment where the reader interacts with the plots in order to develop his or her own understanding.
It should be noted that for disks that are not so extremely optically thick as Arp 220 the effects will vary somewhat. 

\begin{figure}[ht!]
\plotone{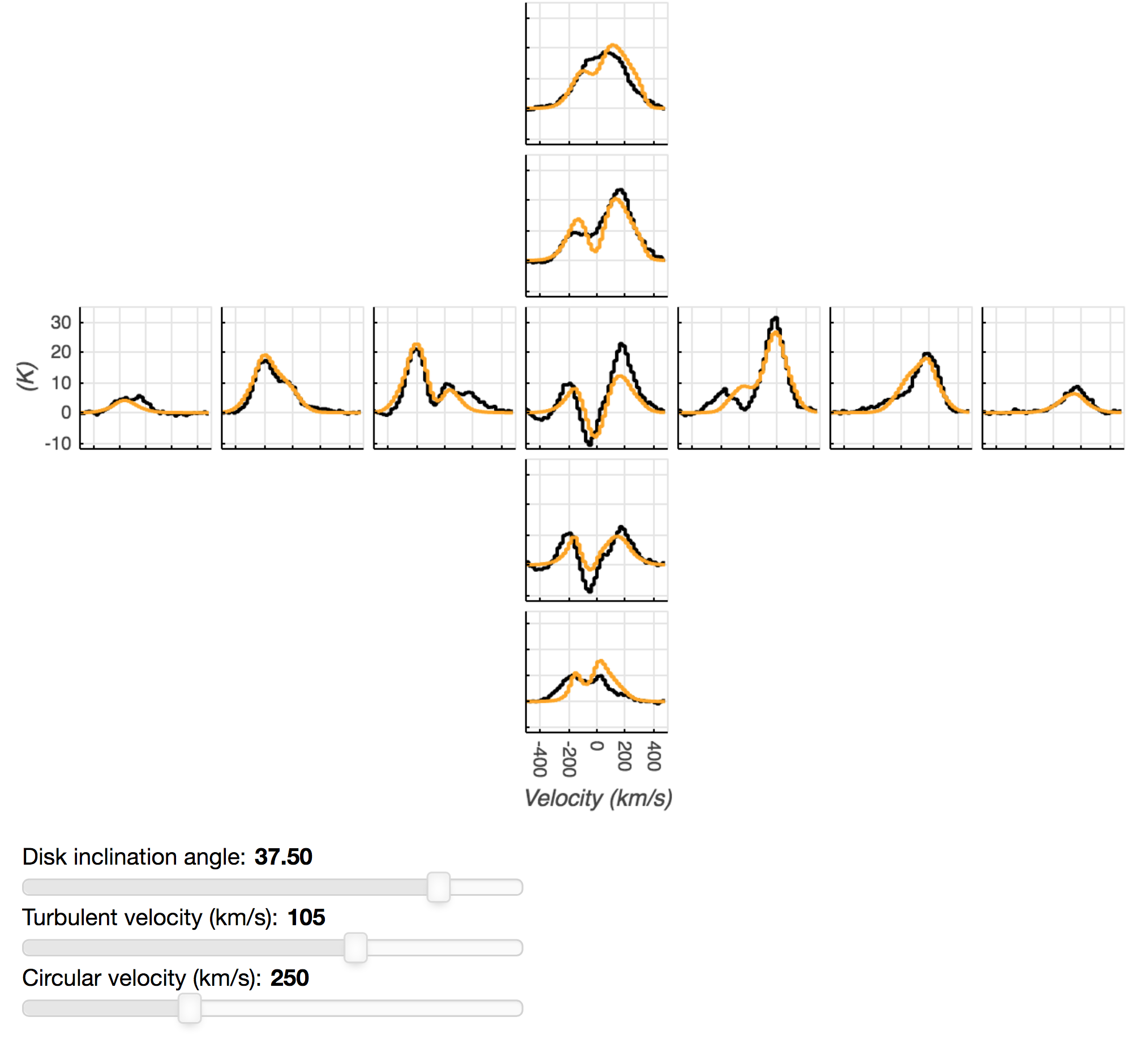}
\caption{Models for the Arp 220 western nucleus in $^{13}$CO J = $4\rightarrow3$ (orange lines) shown along with the observed line profiles at half-beam increments along the disk axis and perpendicular to the disk axis. 1 Jy/beam = 294.5 K. In this figure, any combination of parameters is allowed. Interactive figure available on-line. \label{fig:west_cube}}
\end{figure}

In Figure \ref{fig:west_cube},  a cube of models varying the turbulent velocity, the circular velocity, and inclination angle of the disk is depicted. 
As an example of the degeneracy, the reader can select a turbulent velocity of 95 km/s, a circular velocity of 280 km/s, and an inclination angle of 30$^\circ$ versus a turbulent velocity of 105 km/s, a circular velocity of 250 km/s, and an inclination angle of 37.5$^\circ$.
Note that the profiles, while different, are quite similar for the two sets of disparate conditions. 
The take away is that if the disk is taken to have a different inclination angle different results for the turbulent velocity and circular velocity will be found and vice versa.

\begin{figure}[ht!]
\plotone{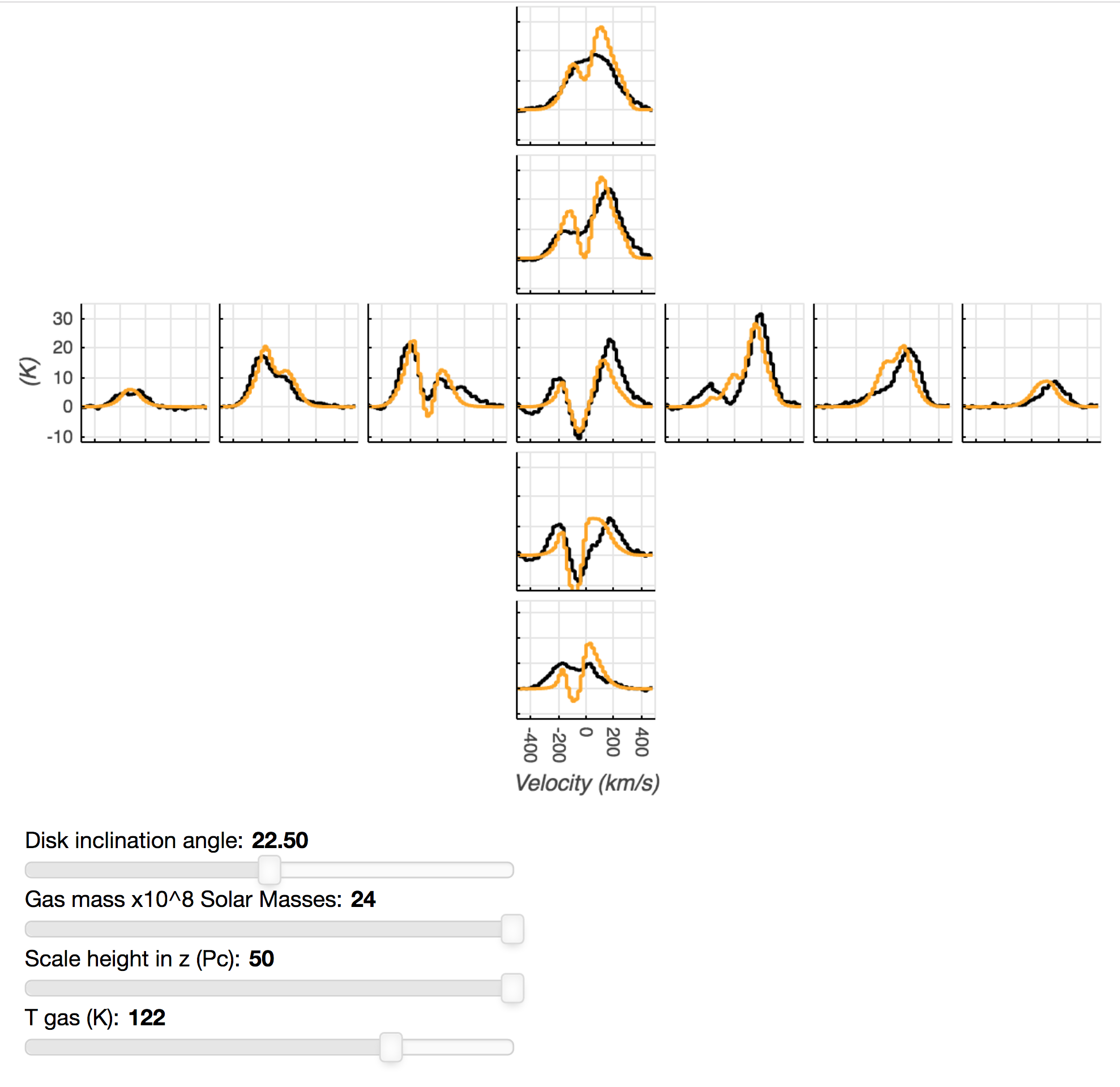}
\caption{Models for the Arp 220 western nucleus in $^{13}$CO J = 4-3 ( orange lines) shown along with the observed line profiles at half-beam increments along the disk axis and perpendicular to the disk axis. 1 Jy/beam = 294.5 K. In this figure, any combination of parameters is allowed. Interactive figure available on-line. \label{fig:west_hyper_cube}}
\end{figure}

In Figure \ref{fig:west_hyper_cube}, models varying the gas temperature,  the thickness of the disk,  total gas mass, and inclination angle of the disk are shown. 
As an example of the degeneracy, the reader can select an inclination angle of 30$^\circ$, a gas mass of 16 $\times 10^8$ M$_\odot$, a scale height in z of 35 pc, and a gas temperature of 112 K versus an inclination angle of 22.5$^\circ$, a gas mass of 4$\times 10^8$ M$_\odot$, a scale height in z of 50 pc, and a gas temperature of 122 K.
Notice how similar the central absorption dip is between these two sets of parameters.
For the models in this paper, the dust temperature is partly constrained by the continuum flux density given a total gas mass, therefore not every parameter combination is plausible.

%% This command is needed to show the entire author+affilation list when
%% the collaboration and author truncation commands are used.  It has to
%% go at the end of the manuscript.
%\allauthors

%% Include this line if you are using the \added, \replaced, \deleted
%% commands to see a summary list of all changes at the end of the article.
%\listofchanges


\begin{thebibliography}{}

\bibitem[Aalto et al.(2007)]{Aalto2007} Aalto, S., Spaans, M., Wiedner, M.~C., \& H{\"u}ttemeister, S.\ 2007, \aap, 464, 193 
\bibitem[Aalto et al.(2015)]{Aalto2015} Aalto, S., Mart{\'\i}n, S., Costagliola, F., et al.\ 2015, \aap, 584, A42.
\bibitem[Aalto et al.(2019)]{2019A&A...627A.147A} Aalto, S., Muller, S., K{\"o}nig, S., et al.\ 2019, \aap, 627, A147
\bibitem[Barcos-Mu{\~n}oz et al.(2015)]{Barcos2015} Barcos-Mu{\~n}oz, L., Leroy, A.~K., Evans, A.~S., et al.\ 2015, \apj, 799, 10 
\bibitem[Barcos-Mu{\~n}oz et al.(2018)]{Barcos2018} Barcos-Mu{\~n}oz, L., Aalto, S., Thompson, T.~A., et al.\ 2018, \apjl, 853, L28
\bibitem[Barger et al.(1998)]{Barger1998} Barger, A.~J., Cowie, L.~L., Sanders, D.~B., et al.\ 1998, \nat, 394, 248 
\bibitem[Barnes \& Hernquist(1991)]{Barnes1991} Barnes, J.~E., \& Hernquist, L.~E.\ 1991, \apjl, 370, L65 
\bibitem[Bolatto et al.(2013)]{Bolatto2013} Bolatto, A.~D., Wolfire, M., \& Leroy, A.~K.\ 2013, \araa, 51, 207.
\bibitem[Brinch \& Hogerheijde(2010)]{Brinch2010} Brinch, C. \& Hogerheijde, M.~R.\ 2010, \aap, 523, A25.
\bibitem[Casey et al.(2014)]{Casey2014} Casey, C.~M., Narayanan, D., \& Cooray, A.\ 2014, \physrep, 541, 45.
\bibitem[Cox et al.(2006)]{Cox2006} Cox, T.~J., Jonsson, P., Primack, J.~R., et al.\ 2006, \mnras, 373, 1013.
\bibitem[Dav{\'e} et al.(2010)]{Dave2010} Dav{\'e}, R., Finlator, K., Oppenheimer, B.~D., et al.\ 2010, \mnras, 404, 1355.
\bibitem[Downes \& Solomon(1998)]{DownesSolomon1998} Downes, D., \& Solomon, P.~M.\ 1998, \apj, 507, 615 
\bibitem[Draine \& Salpeter(1979)]{Draine1979} Draine, B.~T., \& Salpeter, E.~E.\ 1979, \apj, 231, 77 
\bibitem[Engel et al.(2010)]{Engel2010} Engel, H., Tacconi, L.~J., Davies, R.~I., et al.\ 2010, \apj, 724, 233.
\bibitem[Espada et al.(2012)]{Espada2012} Espada, D., Komugi, S., Muller, E., et al.\ 2012, \apj, 760, L25.
\bibitem[Gonz{\'a}lez-Alfonso et al.(2012)]{Gonzalez2012} Gonz{\'a}lez-Alfonso, E., Fischer, J., Graci{\'a}-Carpio, J., et al.\ 2012, \aap, 541, A4
\bibitem[Gonz{\'a}lez-Alfonso et al.(2013)]{Gonzalez2013} Gonz{\'a}lez-Alfonso, E., Fischer, J., Bruderer, S., et al.\ 2013, \aap, 550, A25
\bibitem[Gonz{\'a}lez-Alfonso \& Sakamoto(2019)]{Gonzalez2019} Gonz{\'a}lez-Alfonso, E., \& Sakamoto, K.\ 2019, \apj, 882, 153
\bibitem[Haris et al.(2016)]{Haris2016} Haris, U., Parvathi, V.~S., Gudennavar, S.~B., et al.\ 2016, \aj, 151, 143 
\bibitem[Hughes et al.(1998)]{Hughes1998} Hughes, D.~H., Serjeant, S., Dunlop, J., et al.\ 1998, \nat, 394, 241 
\bibitem[Johnston et al.(2008)]{Johnston2008} Johnston, K.~V., Bullock, J.~S., Sharma, S., et al.\ 2008, \apj, 689, 936 
\bibitem[Koprowski et al.(2016)]{Koprowski2016} Koprowski, M.~P., Dunlop, J.~S., Micha{\l}owski, M.~J., et al.\ 2016, \mnras, 458, 4321.
\bibitem[Lonsdale et al.(2006)]{Lonsdale2006} Lonsdale, C.~J., Diamond, P.~J., Thrall, H., Smith, H.~E., \& Lonsdale, C.~J.\ 2006, \apj, 647, 185 
\bibitem[Mart{\'\i}n et al.(2011)]{Martin2011} Mart{\'\i}n, S., Krips, M., Mart{\'\i}n-Pintado, J., et al.\ 2011, \aap, 527, A36.
\bibitem[Mart{\'\i}n et al.(2016)]{Martin2016} Mart{\'\i}n, S., Aalto, S., Sakamoto, K., et al.\ 2016, \aap, 590, A25.
\bibitem[McMullin et al.(2007)]{CASA} McMullin, J.~P., Waters, B., Schiebel, D., et al.\ 2007, Astronomical Data Analysis Software and Systems XVI, 127.
\bibitem[Men{\'e}ndez-Delmestre et al.(2013)]{Menendez2013} Men{\'e}ndez-Delmestre, K., Blain, A.~W., Swinbank, M., et al.\ 2013, \apj, 767, 151.
\bibitem[Micha{\l}owski et al.(2017)]{Michalowski2017} Micha{\l}owski, M.~J., Dunlop, J.~S., Koprowski, M.~P., et al.\ 2017, \mnras, 469, 492 
\bibitem[Narayanan et al.(2009)]{Narayanan2009} Narayanan, D., Cox, T.~J., Hayward, C.~C., et al.\ 2009, \mnras, 400, 1919.
\bibitem[Narayanan et al.(2015)]{Narayanan2015} Narayanan, D., Turk, M., Feldmann, R., et al.\ 2015, \nat, 525, 496.
\bibitem[Ossenkopf \& Henning(1994)]{Ossenkopf1994} Ossenkopf, V., \& Henning, T.\ 1994, \aap, 291, 943
\bibitem[Rangwala et al.(2011)]{Rangwala2011} Rangwala, N., Maloney, P.~R., Glenn, J., et al.\ 2011, \apj, 743, 94.
\bibitem[Rangwala et al.(2015)]{Rangwala2015} Rangwala, N., Maloney, P., Wilson, C., et al.\ 2015, IAU General Assembly 22, 2255618.
\bibitem[Sakamoto et al.(1999)]{Sakamoto1999} Sakamoto, K., Scoville, N.~Z., Yun, M.~S., et al.\ 1999, \apj, 514, 68 
\bibitem[Sakamoto et al.(2008)]{Sakamoto2008} Sakamoto, K., Wang, J., Wiedner, M.~C., et al.\ 2008, \apj, 684, 957
\bibitem[Sakamoto et al.(2009)]{Sakamoto2009} Sakamoto, K., Aalto, S., Wilner, D.~J., et al.\ 2009, \apj, 700, L104.
\bibitem[Sakamoto et al.(2010)]{Sakamoto2010} Sakamoto, K., Aalto, S., Evans, A.~S., Wiedner, M.~C., \& Wilner, D.~J.\ 2010, \apjl, 725, L228 
\bibitem[Sakamoto et al.(2017)]{Sakamoto2017} Sakamoto, K., Aalto, S., Barcos-Mu{\~n}oz, L., et al.\ 2017, \apj, 849, 14.
\bibitem[Sargent et al.(2012)]{Sargent2012} Sargent, M.~T., B{\'e}thermin, M., Daddi, E., et al.\ 2012, \apj, 747, L31.
\bibitem[Savage \& Sembach(1996)]{Savage1996} Savage, B.~D., \& Sembach, K.~R.\ 1996, \araa, 34, 279 
\bibitem[Scoville et al.(2017)]{Scoville2017} Scoville, N., Murchikova, L., Walter, F., et al.\ 2017, \apj, 836, 66
\bibitem[Scoville et al.(2017)]{Scoville2017b} Scoville, N., Lee, N., Vanden Bout, P., et al.\ 2017, \apj, 837, 150.
\bibitem[Schilke et al.(1997)]{Schilke1997} Schilke, P., Walmsley, C.~M., Pineau des Forets, G., \& Flower, D.~R.\ 1997, \aap, 321, 293 
\bibitem[Solomon et al.(1992)]{Solomon1992} Solomon, P.~M., Downes, D., \& Radford, S.~J.~E.\ 1992, \apjl, 398, L29 
\bibitem[Sparre \& Springel(2016)]{Sparre2016} Sparre, M., \& Springel, V.\ 2016, \mnras, 462, 2418.
\bibitem[Taylor et al.(2001)]{Taylor2001} Taylor, C.~L., Walter, F., \& Yun, M.~S.\ 2001, \apjl, 562, L43 
\bibitem[Teyssier et al.(2010)]{Teyssier2010} Teyssier, R., Chapon, D., \& Bournaud, F.\ 2010, \apjl, 720, L149 
\bibitem[Tunnard et al.(2015)]{Tunnard2015} Tunnard, R., Greve, T.~R., Garcia-Burillo, S., et al.\ 2015, \apj, 800, 25.
\bibitem[Wilson et al.(2014)]{Wilson2014} Wilson, C.~D., Rangwala, N., Glenn, J., et al.\ 2014, \apj, 789, L36.
\bibitem[Yoast-Hull et al.(2017)]{Yoast2017} Yoast-Hull, T.~M., Gallagher, J.~S., Aalto, S., et al.\ 2017, \mnras, 469, L89

 


\end{thebibliography}
\end{document}